\documentclass[aps,pra,twocolumn,showpacs]{revtex4-1}
\usepackage{amsbsy,latexsym,amsmath}
\usepackage{amsfonts}
\usepackage{amssymb}
\usepackage[mathscr]{eucal}
\usepackage{epsfig,graphics,graphicx}

\newcommand{\ket}[1]{\vert#1\rangle}
\newcommand{\bra}[1]{\langle#1\vert}
\newcommand{\braket}[2]{\langle#1\vert#2\rangle}

\newcommand{\expo}[1]{\mathrm{e}^{#1}}
\newcommand{\abs}[1]{\left|{#1}\right|}
\renewcommand{\vec}[1]{\mbox{\boldmath$#1$}}
\newcommand{\scrvec}[1]{\mbox{\boldmath\scriptsize$#1$}}

\begin{document}
\title{Optimal parameter estimation with a fixed rate of abstention}
\author{B. Gendra, E. Ronco-Bonvehi, J. Calsamiglia, R. Mu\~{n}oz-Tapia, and E. Bagan}
\address{F\'{i}sica Te\`{o}rica: Informaci\'{o} i Fen\`{o}mens Qu\`antics, Departament de F\'{\i}sica, Universitat Aut\`{o}noma de Barcelona, 08193 Bellaterra (Barcelona), Spain}

\begin{abstract}
The problems of optimally estimating a phase, a direction, and the orientation of a Cartesian  frame (or trihedron) with general pure states 
are addressed. Special emphasis is put on estimation schemes that allow for inconclusive answers or abstention. It is shown that such schemes enable drastic improvements, up to the extent of attaining the Heisenberg limit in some cases, and the required amount of abstention is quantified. 
A general mathematical framework to deal with the asymptotic limit of many qubits or large angular momentum is introduced and used to obtain analytical results for all the relevant cases under consideration. Parameter estimation with abstention is also formulated as a semidefinite programming problem, for which very efficient numerical optimization techniques~exist.

%

\end{abstract}

\maketitle

\section{Introduction}

State identification and state estimation are fundamental and highly non-trivial tasks in quantum information. The main difficulty lies in the fact that quantum measurements provide partial information about the state of a quantum system, and only when a large number $N$ of identically prepared copies of  such a system is available to an experimentalist can she attempt to accomplish a successful identification or a faithful estimation.

In standard protocols the experimentalist is expected to produce a conclusive answer (maybe not right or accurate enough), based on the outcomes of her measurements, at each run of the experiment. To assess the overall performance of the protocol an average cost function or figure of merit is computed, e.g., the minimum probability~of misidentification, or the fidelity, $F$, between estimate and true state. In this context, many results have been obtained over the last years in a large variety of settings~\mbox{\cite{holevo,paris-rehacek,massar-popescu,hradil,derka-buzek,bruss,fischer,bagan-local,bagan-mixed,bagan-optimal,blume,steinberg,gross,helstrom,bergourev}}.

A new class of protocols has recently emerged as a viable alternative in situations where the approach discussed above fails to achieve the minimum standard of performance required for some specific task, but some number of inconclusive responses, or abstentions, is affordable. This can be seen as a particular instance of post-selection. Examples of such protocols can be found in state discrimination~\cite{unambiguous,chefles2,zhang,fiurasek,eldar,hayashi,sugimoto,bagan-q}, where some fixed rate $Q$ of inconclusive outcomes can raise the probability of success significantly (or, e.g.,  lower the probability of error even down to zero, as in unambiguous discrimination~\cite{unambiguous}), and also in state estimation~\cite{massar-popescu2}, where abstention is shown to reduce the negative impact of noisy detectors~\cite{gendra}.  


In this paper we consider the natural extension of this approach to quantum parameter estimation with~pure states of~$N$ qubits (or Rydberg atomic states of total angular momentum $N/2$). More precisely, we deal with an infinite covariant family of such~states, paramatrised by some continuous variables, and we aim to estimate the values of these variables  for a given sample state by performing suitable measurements on it. 
We already presented in \cite{qmet} the paradigmatic instance of phase estimation. Here, we will provide the details missing in~\cite{qmet} and 
will also address the problem of estimation of spatial directions, which~we assume encoded in a given $N$-spin or angular momentum state. In particular, we focus on two problems, that of a single direction and that of three mutually orthogonal directions (trihedron or Cartesian frame); the later will be referred to as frame estimation for brevity.


%
%

The use of abstention in the context of estimation was previously considered in~\cite{fiurasek2}, where the author dealt with qudit pure state estimation from a pair of conjugate qudits, and also with estimation of an equatorial qubit state from $N$ independent and uncorrelated copies of the~state (phase estimation). At variance with our approach, no lower limit on the acceptance rate (i.e.,~$\bar Q\equiv 1-Q$) was imposed. In the phase estimation example he considered, the increase in fidelity was achieved only at the cost of imposing acceptance rates that vanish exponentially as~$N$ goes to infinity.

The approach of Ref.~\cite{gendra} to multiple-copy qubit state estimation with non-ideal measurements is another example of parameter estimation with abstention. In this case the covariant family of states is the set~$\{\rho^{\otimes N}(r\vec n)\}_{\scrvec n \in{\mathbb S}^2}$, where $r\vec n$ is the Bloch vector of~$\rho(r\vec n)$ and the unit vector~$\vec n$ parametrizes the family. The parameter space is thus the unit  $2$-sphere~${\mathbb S}^2$; in this example, the purity $r$ takes the same fixed value for the entire family of states. It is shown in~\cite{gendra} that abstention leads to a significant increase in the average fidelity for small samples, but for asymptotically large $N$ the fidelity enhancement is modest, and besides, it requires an exponentially vanishing acceptance rate. 
%
%
To summarize, in the cases previously considered in the literature, abstention has limited impact on parameter estimation with asymptotically large samples unless the experimentalist abstains from producing an estimate most of the time.


In~\cite{qmet} we showed that the situation changes dramatically for phase estimation if one allows for more general covariant families. Here we will show that this is also the case for the two new problems at hand, namely, for
single direction and frame estimation with pure states of~$N$ qubits. 
For phase estimation the covariant family we are referring to is the set of states of the form~\mbox{$\{|\Psi(\theta)\rangle=U(\theta)|\Psi_0\rangle\}_{\theta\in[0,2\pi)}$}, where~$U(\theta)$~stands for the unitary transformation~$U(\theta)|j\rangle={\rm e}^{i\theta j}|j\rangle$, $|\Psi_0\rangle$ is a fiducial state, which in the eigenbasis of $U(\theta)$ can be written as $|\Psi_0\rangle=\sum_{j=0}^n c_j |j\rangle\in ({\mathbb C}^{2})^{\otimes n}$, and the number of qubits is~$N=n$.  The components $c_j$ are given arbitrary coefficients, subject to 
the normalization condition~$\sum_{j=0}^n|c_j|^2=1$.
For direction estimation we consider instead~\mbox{$\{|\Psi(\vec n)\rangle =U_{\scrvec n}|\Psi_0\rangle\}_{\scrvec n \in {\mathbb S}^2}$},  where $U_{\scrvec n}$ stands for the unitary representation of the rotation that takes $\vec z$ (the unit vector in the $z$-axis; likewise, $\vec x$ and $\vec y$ stand for the other two unit vectors) into $\vec n$. The fiducial state is now given by~\mbox{$|\Psi_0\rangle=\sum_{j=0}^{n}c_j|j,0\rangle$}, which may be thought of as pointing along the $z$-axis  (in the sense that it is invariant under rotations about that axis),~$N=2n$, and we use the standard notation $|j,m\rangle$ for the total angular momentum eigenstates. The choice~$m=0$ is both for simplicity and also because the optimal state for direction encoding is known to have null total magnetic number \cite{bagan-optimal}, however the method can be extended to any $m$. 
More general states, i.e., those that are not eigenstates of $J_z$, do not fit into our pure state framework~\cite{abst-mixed}, since for the sake of direction estimation the subset $\{{\rm e}^{-i \gamma J_z}|\Psi_0\rangle\}_{\gamma\in[0,2\pi)}$ that encodes $\vec z$ is equivalent to the mixed state $\rho_0=\int_0^{2\pi}(d\gamma/2\pi){\rm e}^{-i \gamma J_z} |\Psi_0\rangle\langle\Psi_0|{\rm e}^{i \gamma J_z}$. 

For frame estimation, the relevant family of states is~$\left\{|\Psi(g)\rangle=U(g)|\Psi_0\rangle\right\}_{g\in {\mathbb S}^3}$, where $g$ stands for the three Euler angles: $g=(\alpha,\beta,\gamma)$. They specify the rotation that takes the axes $x$, $y$ and $z$ into those of the Cartesian frame we wish to estimate, with unit vectors $(\vec n^1,\vec n^2,\vec n^3)$. It can be shown that optimality requires a fiducial state of the form
$
\sum_{j=0}^{n} c_j \,(\sum_{m=-j}^j|j,m,\alpha_m\rangle)/\sqrt{2j+1}$,
where~$N=2n$ and the third quantum number in the ket,~$\alpha_m$, labels the degeneracy of the representation of angular momentum $j$.
Except for the representation of highest angular momentum, $j=n$,  for each~$j<n$ we have (maximally) entangled the magnetic number $m$ with the degeneracy number $\alpha_m$. This entanglement with `ancillary' degrees of freedom that are invariant under the action of the group is responsible for an important enhancement in the estimation precision \cite{framesopt}.
We note in passing that this degeneracy is known to be useless for single direction estimation, and thus we dropped the corresponding label there.  Indeed, following the symmetry argument used at the end of the previous paragraph, any entanglement between magnetic number and degeneracy labels would in effect turn into an incoherent sum on subspaces of different $m$ values, which is clearly suboptimal.


From a formal point of view, it will be seen that the optimization of the frame estimation protocol for this family of states is equivalent to that of phases for large $N$. Thus, we find it more interesting to ignore the degeneracy of the representations and consider instead the family generated by the fiducial state
$|\Psi_0\rangle=\sum_{j=0}^{n} c_j |j,j\rangle$. States of this form could be produced if, e.g., a hydrogen atom in a Rydberg state of total angular momentum up to $n$ is used instead of $N$ spins~\cite{peresframes}. In this scenario, the optimal encoding state for a Cartesian frame is known to belong to this family, but it does not lead to a Heisenberg scaling precision. 
 Also in this case, the method we will introduce can be applied to more general pure states.

For all these estimation problems, a finite acceptance rate $\bar Q$ suffices to lower the coefficient of the leading order in the asymptotic expansion of the average error in inverse powers of $N$. If an exponentially vanishing acceptance rate is affordable, the leading order in this expansion becomes~$1/N^2$, thus attaining the Heisenberg limit, except for frame estimation with Rydberg states.
It will be shown that
 the effect of abstention can be understood in terms of a probabilistic map from the original family to a  better one (closer to optimal), $\{\tilde\Psi(\theta)\}_{\theta\in[0,2\pi)}$ (or~$\{\tilde\Psi(\vec n)\}_{\scrvec n\in{\mathbb S}^2}$, etc.), which fails with probability $Q$.
%

Last but not least, here we present a general technique to obtain the asymptotic form of pure state parameter estimation problems, with or without abstention, that is interesting on its own. The main idea is that the components of $|\tilde\Psi_0\rangle$ can be viewed as a discretization of some continuous function $\varphi(t)$ on the unit interval $[0,1]$,
and 
likewise, the problem of maximizing the fidelity over  those components can be viewed (see below) as a discretization of a constrained variational problem for $\varphi(t)$. The solution of the latter problem gives the asymptotic expression of the fidelity for the former one. This solution can be worked out analytically for many physically relevant settings.
%
%
%
%
For finite $N$ the estimation can be formulated as a semidefinite programming (SDP) problem, and hence solved numerically with very high efficiency~\cite{SDP}.



\section{General framework}

The problems of phase, direction and frame estimation described above can be treated in a unified framework by writing $U(\theta)=U(g)$,  $|\Psi(\theta)\rangle=|\Psi(g)\rangle$, where~\mbox{$g\in{\mathbb S}^1$}; and~$U_{\scrvec n}=U(g)$, $|\Psi(\vec n)\rangle=|\Psi(g)\rangle$, where~$g\in{\mathbb S}^2$.  Since the magnetic number is fixed to zero ($j$) for direction (frame
) estimation, we also drop this quantum number and write~$|j,0\rangle\equiv |j\rangle$  ($|j,j\rangle\equiv |j\rangle$). Then, for the three problems we have a family of states $\{|\Psi(g)\rangle=U(g)|\Psi_0\rangle\}_{g \in{\mathbb S}^d}$, where~$d=1$, $2$, $3$ for phase, direction, and frame estimation, respectively. 
As already mentioned above, in direction (frame) estimation the fiducial state~$|\Psi_0\rangle$ can be thought of as encoding the unit vector~$\vec{z}$ [the cartesian frame $(\vec{x},\vec{y},\vec{z})$]. Similarly, in phase estimation, $|\Psi_0\rangle$ can be interpreted as encoding the reference unit vector $\vec{x}$ (to which we assign a zero phase), and~$U(g)$ as a rotation of (Euler) angle $\alpha=\theta$ around the~$z$-axis~%
\footnote{Note that from this point of view the the label~\mbox{$j=0,\ldots, n$} would correspond to the magnetic quantum number taking values $m=-n/2,\dots, n/2$.}%
.
Hence, in this unified framework, we can define a cost function in terms of the (quadratic) error per axis is: i.e.~\mbox{$e_1(g,g_\chi)=|\vec n-\vec n_\chi|^2$} for phase and direction estimation,  and  the total error~$e_3(g,g_\chi)=\sum_{a=1}^3|\vec n^a-\vec n^a_\chi|^2$ for frame estimation. In these expressions, the subscript~$\chi$ specifies that the estimate is based on the outcome $\chi$ of a generalized measurement that will be introduced below. These errors are related to the `relative rotation' $U^\dagger(g_\chi)U(g)=U(g_\chi^{-1}g)$ through
\begin{eqnarray}
e_1(g,g_\chi)&=&2-2\langle1,0|U(g_\chi^{-1}g)|1,0\rangle ,
\label{ebc04.06.13-1a}
\\
e_3(g,g_\chi)&=&6-2\sum_{m=-1}^1\langle1,m|U(g_\chi^{-1}g)|1,m\rangle
,
\label{ebc04.06.13-1b}
\end{eqnarray}
where we recognize the sum in~(\ref{ebc04.06.13-1b}) as the character of~$U(g_\chi^{-1}g)$  in the $j=1$ representation. Note that~$0\le e_1(g,g_\chi)\le 4$, and $0\le e_3(g,g_\chi)\le 8$ (we can at most get two axes completely wrong since we assume right-handed Cartesian frames). As a figure of merit, the fidelity
$f(g,g_\chi)=(1+\vec n\cdot\vec n_\chi)/2=1-e_1(g,g_\chi)/4$ is most commonly used in phase and direction estimation. One has $0\le f(g,g_\chi)\le 1$, where~$1$ corresponds to perfect estimation. For frame estimation one can also define a fidelity with the same range of values as~$f(g,g_\chi)=1-e_3(g,g_\chi)/8$. These fidelities are also trivial functions of the relative rotation $U(g^{-1}_\chi g)$ in the $j=1$ representation through  Eqs.~(\ref{ebc04.06.13-1a}) and~(\ref{ebc04.06.13-1b}).


The generalized measurements we are interested in are  characterized mathematically by a positive operator valued measure (POVM) $\Pi=\{\Pi_\chi\}_{\chi\in {C}}\cup\{\Pi_0\}$, where $\Pi_0$ and each $\Pi_\chi$ are non-negative operators that add up to the identity, i.e., $\Pi_0+\sum_{\chi\in {C}}\Pi_\chi=\openone$, $C$ is the set of conclusive outcomes (from which an estimate is proposed) and $\Pi_0$ outputs `abstention'. The probability of such abstention taking place is
\begin{equation}\label{def Q}
Q=\int_{{\mathbb S}^d} d^dg\, \langle\Psi(g)|\Pi_0|\Psi(g)\rangle,
\end{equation}
and $\bar Q=1-Q$ is the acceptance probability (rate at which we provide definite estimates).
The average fidelity for this rate of abstention is
\begin{equation}
F(Q)={1\over \bar Q}\sum_\chi\int_{{\mathbb S}^d} d^dg\,f(g,g_\chi) \,\langle\Psi(g)|\Pi_\chi|\Psi(g)\rangle.
\label{eq:fidq}
\end{equation}
In Eqs.~(\ref{def Q}) and~(\ref{eq:fidq}), $d^dg$ stands for the (normalized) `volume' elements $dg=d\alpha/(2\pi)$ (for $d=1$), $d^2g=\sin\beta\, d\alpha\, d\beta/(4\pi)$, and $d^3g=\sin\beta\, d\alpha\, d\beta\,d\gamma /(8\pi^2)$. They are invariant measures  on ${\mathbb S}^d$, i.e., $d^d(gg')=d^dg$.

%
%
Estimation with abstention can be reduced to a standard estimation problem (without abstention) by simply introducing the new POVM $\tilde\Pi$, with elements given by
\begin{equation}
\tilde\Pi_\chi\equiv\left(\openone-\Pi_0\right)^{-1/2} \Pi_\chi  \left(\openone-\Pi_0\right)^{-1/2} ,
\end{equation}
and the new family of (normalized) states
\begin{equation}
\left\{|\tilde\Psi(g)\rangle\equiv{\left(\openone-\Pi_0\right)^{1/2} \over\bar Q^{1/2}}|\Psi(g)\rangle\right\}_{g\in{\mathbb S}^d}  .
\end{equation}
With these two definitions we can write the fidelity as
\begin{equation}
F(\Pi_0)=\sum_\chi\int_{{\mathbb S}^d} d^dg\, f(g,g_\chi)\,\langle\tilde\Psi(g)|\tilde\Pi_\chi|\tilde\Psi(g)\rangle,
\label{eq:fid}
\end{equation}
where we emphasize that this expression depends on the choice of $\Pi_0$. 
This expression also brings forward an interpretation of the role of abstention in this optimization problem that we will use throughout the paper:  each  initial state  $|\Psi(g)\rangle$ is transformed into a new $|\tilde\Psi(g)\rangle$ that encodes the unknown parameter(s) $g$  in a more efficient way.  
This map improves the estimation precision by effectively increasing the distinguishability between the signal states, therefore it can only be implemented in a probabilistic fashion (it succeeds with probability~$\bar Q$). This stochastic map is fully specified  by the optimal choice of~$\Pi_0$:
\begin{equation}\label{max F}
F(Q)=\max_{\Pi_0\,:\,\mbox{\footnotesize Eq.~(\protect\ref{def Q})}} F(\Pi_0)  .
\end{equation}
Although this may seem a difficult optimization problem, a huge simplification arises because of the covariance of the family of states. Already from Eqs.~(\ref{def Q}) and \eqref{eq:fidq} one can 
easily  see  that the optimal POVM can be chosen to be covariant under the set of unitaries~$\{U(g)\}_{g\in{\mathbb S}^d}$. In particular this means that $\Pi_0$  can be taken invariant under the corresponding unitary group. For $d=1$ this is just the group $U(1)$. For $d=2$ ($d=3$) the integral 
over the $2$-sphere ($3$-sphere) can be turned into (is) a $SU(2)$ group integral. Thus, Shur's lemma can be applied to all the cases, which results in $\Pi_0$ being proportional to the identity on each irreducible block: $\Pi_0=\sum_j f_j |j\rangle\langle j|$ (phase estimation), \mbox{$\Pi_0=\sum_j f_j\,\sum_m|j,m\rangle\langle j,m|\equiv\sum_j f_j\openone_j$} (direction and frame estimation). Hence, the maximization in Eq.~(\ref{max F}) is over $\{f_j\,:\, 0\le f_j\le 1\}_{j=0}^n$. Note that the transformed set of states $\{|\tilde\Psi(g)\rangle\}_{g\in{\mathbb S}^d}$ is also a covariant family, just as the original one. The corresponding reference state is 
\begin{equation}\label{transformed}
|\tilde\Psi_0\rangle=\sum_{j=0}^n{ c_j\sqrt{\bar f_j}\over \sqrt{\bar Q}}|j\rangle=
\sum_{j=0}^n\xi_j|j\rangle \equiv\ket{\xi} ,
\end{equation} 
where $\bar f_j\equiv 1-f_j$. From  Eq.~(\ref{def Q}), and using Shur's lemma, we find
\begin{equation}
Q=\sum_{j=0}^n |c_j|^2f_j=1-\sum_{j=0}^n |c_j|^2\bar f_j  .
\end{equation}
Thus, $|\tilde\Psi_0\rangle=\ket{\xi}$ is a normalized state, as it should, i.e.,~$\sum_j|\xi_j|^2=1$.
Since the transformed states are still covariant, we can choose the POVM~$\tilde \Pi$ to be the well known continuous and covariant POVM for each of the problems at hand \cite{holevo,bagan-optimal}: $\{\tilde\Pi_{g}=U(g)|\Phi_d\rangle\langle\Phi_d| U^\dagger(g)\}_{g\in{\mathbb S}^d}$, where the unnormalized state~$|\Phi_d\rangle$ is given by
\begin{equation}\label{seeds}
|\Phi_1\rangle=\sum_{j=0}^n|j\rangle;\quad |\Phi_{2,3}\rangle=\sum_{j=0}^n\sqrt{2j+1}|j\rangle.
\end{equation}
Note that $g$ plays the role of $\chi$, i.e., $g$ specifies the different outcomes of the measurement.
Hereafter in this paper, it is assumed that the states have non-negative coefficients  $c_{j}\geq 0$ (and hence $\xi_{j}\geq 0$). This is a valid assumption since any phases present in the coefficients $c_{j}$ (or $\xi_{j}$) can be absorbed by the above POVM's. 
This result makes the calculation of the fidelity $F(\Pi_0)$ straightforward:
%
\begin{equation}\label{fidelity-general-3}
F(\Pi_0)={1\over2}+ {1\over2}\bra{\xi} \mathsf{M}  \ket{\xi},
\end{equation}
where in the canonical basis, $\{\ket{j}\}_{j=0}^{n}$,
$\mathsf{M}$ is a real matrix of tridiagonal form
\begin{equation}\label{matrix}
\mathsf{M}=\begin{pmatrix}
h^d_0&a^d_{1\phantom{-1}}&
         &        &       \cr
                    a^d_{1\phantom{-1}}&\ddots&\ddots
    &\phantom{\ddots} \raisebox{2.0ex}[1.5ex][0ex]{\LARGE
0}\hspace{-.5cm}       &       \cr
                         &\ddots&h^d_{n-2}    &a^d_{n\!-\!1\phantom{-1}}  &       \cr
                         & \phantom{\ddots}    &a^d_{n\!-\!1\phantom{-1}}&h^d_{n-1}&a^d_{n\phantom{-1}}
\cr \hspace{.5cm} \raisebox{2.0ex}[1.5ex][0ex]{\LARGE
0}\hspace{-.5cm} &&\phantom{\ddots}&a^d_{n\phantom{-1}}&h^d_n
\end{pmatrix} ,
\end{equation}
with
\begin{eqnarray}
&\displaystyle a^1_j={1\over2}; \quad a^2_j={j\over\sqrt{4j^2-1}},\quad  a^3_j={1\over2}\sqrt{2j+1\over2j+3},&\nonumber\\
&\displaystyle h^1_j=h^2_j=0,\qquad h^3_j=-{1\over 2(j+1)},&
\end{eqnarray}
where we recall that the superscripts $1$, $2$ and $3$ refer to phase, direction and frame estimation, respectively. 

At this point one can easily check our statement in the introduction that frame estimation with the family generated by the fiducial state~$
\sum_{j=0}^{n} c_j \,(\sum_{m=-j}^j|j,m,\alpha_m\rangle)/\sqrt{2j+1}$ is formally equivalent to phase estimation for large $n$.  For this family, the diagonal entries of  
the matrix $\mathsf{M}$ are zero with the  exception of $h_0=-1/2$ and $h_n=-1/(2n+2)$, whereas the off-diagonal ones are $a_j=1/2$, for~$0\le j\le n-1$ and~$a_n=1/(2\sqrt{2n+1})$. Thus, except for four entries,~$\mathsf M$ is the same for phase and frame estimation. For very large $n$, this finite differences have no effect at leading order and the asymptotic result we will obtain for phases also hold for frames when the degeneracy of the representations is used in the encoding.

Here, we have given the explicit form of~$\mathsf M$ for the particular fiducial states under study . However, it is worth noting that for general states the matrix~$\mathsf M$ will always have a tridiagonal structure and hence the methods that we use readily apply.   As shown in~\cite{bagan-optimal,frames}, this structure is a generic feature that stems from the fact that the fidelity $f(g,g_\chi)$ 
is a linear function of $\langle1,m|U(g_\chi^{-1}g)|1,m'\rangle$ ($j=1$ representation). Its appearance in the integrant of~\eqref{eq:fid} enforces selection rules that prevent the presence~of other off-diagonal elements in $\mathsf M$.

The maximization over~$\{f_j\}$ of~(\ref{fidelity-general-3}) can be turned into a maximization over the transformed states $\ket{\xi}$, namely:
\begin{equation}\label{Delta AMA}
\Delta\equiv\max_{\ket{\xi}} \bra{\xi} \mathsf{M}  \ket{\xi}, 
\end{equation}
subject to the constraints
\begin{eqnarray}\label{norm}
&\;\;\;&\braket{\xi}{\xi}=\sum_{j=0}^{n}\xi_j^2=1,\\
&&\xi_j=\sqrt{\bar f_{j}\over{\bar Q}}c_{j} \leq{ c_j\over\sqrt{\bar Q}}\equiv\lambda c_j,\quad \lambda\ge 1.
\label{sdp}
\end{eqnarray}
Then, the maximum fidelity for a given rate of abstention~$Q$ is~$F(Q)=(1+\Delta)/2$. 

For large enough abstention rates (i.e. large enough values of~$\lambda$) the constraint \eqref{sdp} has no effect (provided all components~$c_j$ are different from zero) and  $\Delta$ becomes the maximum eigenvalue of the matrix $\mathsf M$. In this case, $F(Q\to1)=F^{*}$ is the maximum fidelity that can be achieved by optimizing the components of the fiducial state; these are given by the corresponding eigenvector~$\ket{\xi^{*}}$ of~$\mathsf M$. The resulting fiducial state  thus generates the optimal signal states $|\Psi(g)\rangle$
. From \eqref{sdp} it is straightforward to obtain the critical acceptance rate 
\begin{equation}
\bar{Q}^{*}=\min_{j} \frac{c_{j}^{2}}{{\xi_{j}^{*}}^{2}}.
\end{equation}
That is, for abstention rates such that~\mbox{$Q\ge Q^{*}= 1- \bar{Q}^{*}$} the fidelity attains its absolute maximum value $F^{*}$ (and higher rates cannot improve the estimation quality).
In the other extreme, when no abstention is allowed \mbox{($Q=0$)}, the solution is determined by the constraints~$\xi_j= c_j$ (no maximization is possible), and~\mbox{$\Delta=\langle c|{\mathsf M}|c\rangle$}.

For intermediate values of $Q\in(0,Q^{*})$ the problem becomes more tricky. 
For moderate values of $n$ one can use standard non-linear optimization packages to solve the above constrained convex optimization problem, Eqs.~(\ref{Delta AMA})--(\ref{sdp}). This can also be easily cast as a semidefinite programming (SDP) problem. The SDP approach is efficient and, furthermore,  provides rigorous bounds on the precision of the solution. 
One simply linearizes~these equations by introducing a SDP (positive operator) variable $\mathsf B$ to play the role of $\ket{\xi}\!\bra{\xi}$. The SDP form of Eqs.~(\ref{Delta AMA}) and~(\ref{sdp}) is then
\begin{equation}\label{Delta AMA sdp}
\Delta\equiv\max_{\mathsf B}  \,{\rm tr}\left(\mathsf{M}  \mathsf{B}\right),
\end{equation}
subject to the constraints
\begin{eqnarray}\label{sdp 2}
&\;\;\;&{\rm tr}\, \mathsf{B}=1, \;\;\; \mathsf{B}\geq 0, \nonumber\\
&& \mathsf{B}_{jj}\leq{ |c_j|^2\over \bar Q}\equiv\lambda^2 |c_j|^2,\quad \lambda\ge 1.
\end{eqnarray}
One can easily prove that the optimal $\mathsf B$ for this problem must necessarily have rank one:  since all the entries of~$\mathsf M$ are non-negative, ${\rm tr}({\mathsf M}{\mathsf B})$ increases with increasing values of the off-diagonal entries $\mathsf B_{i,i+1}$.  Their maximum value consistent with positivity is given by rank one matrices.
Therefore, the optimal $\mathsf B$ is of the form $\ket{\xi}\!\bra{\xi}$ and the SDP solution provides in turn a solution of Eqs.~(\ref{Delta AMA})--(\ref{sdp}). 

However, as advertised earlier, the main focus of this work is on the regime of asymptotically large $n$ and, in particular, on presenting an approach that enables obtaining analytical expressions in this regime, thus complementing the SDP analysis. We will first introduce and discuss in some detail the approach for phase estimation. The generalization to direction and frame estimation will be discussed afterwards.

\section{Asymptotic regime: phase estimation}\label{Sec-AsRePhEs}

Here we consider the problem of phase estimation, for which $\bra{\xi}{\mathsf M}\ket{\xi}$ 
can be cast as
\begin{equation}
\bra{\xi}{\mathsf M}\ket{\xi}= \sum_{j=0}^{n-1} \xi_j\xi_{j+1} .
\label{eq:phaseM}
\end{equation}
This expression can be easily rewritten as
\begin{equation}
\bra{\xi}{\mathsf M}\ket{\xi} =
1-\frac{1}{2}\left[\sum_{j=0}^{n-1}\left(\xi_{j+1}-\xi_j\right)^2+\xi_0^2+\xi_{n}^2\right] \!,
\end{equation}
where the first term (unity) results from using the normalization condition~(\ref{norm}). 
Instead of maximizing this expression, we will equivalently minimize $S\equiv 1-\bra{\xi}{\mathsf M}\ket{\xi}$. A slight difficulty arises here because of the inequality constraints in~(\ref{sdp}). To deal with them we need to use the so called Karush-Kuhn-Tucker  (KKT) conditions (see e.g.,~\cite{KKT}), which are a generalization of the Lagrange method. We first have to introduce a multiplier for each constraint: $b^2/2$; $s_j$, $j=0,\dots,n$; much in the same way as the Lagrange method requires. Hence, we will find the local minima of
\begin{eqnarray}
S&=&
\frac{1}{2}\left[\sum_{j=0}^{n-1}\left(\xi_{j+1}-\xi_j\right)^2+\xi_0^2+\xi_{n}^2\right]\nonumber\\
&-&{b^2\over2}\left(\sum_{j=0}^n\xi_j^2-1\right)+\sum_{j=0}^n s_j\left(\xi_j-\lambda c_j\right) .
\label{eq:action}
\end{eqnarray}
Besides the constraints specified in~(\ref{sdp}), which are referred to as
{\em primal feasibility} conditions,
%
we also need to impose the so called
{\em dual feasibility} conditions,
\begin{equation}
s_j\ge0,\quad j=0,1,\dots, n,
\label{eq:kktcond2}
\end{equation}
and, finally,
\begin{equation}
 s_j\left(\xi_j-\lambda c_j\right)=0 ,\quad j=0,1,\dots, n,
 \label{eq:kktcond3}
\end{equation}
known as  {\em complementary slackness} conditions. 

Rather than attempting to solve this system of conditions for arbitrary $n$, which appears to be 
a difficult task, we will take  $n$ to be asymptotically large and reframe the minimization above as a variational problem for a continuous function $\varphi(t)$ in the unit interval~$[0,1]$. To do so, we proceed as follows:
we first note that as $n$ goes to infinity $j/n$ approaches a continuous real variable~$t$. So, we define
\begin{equation}
0\le t\equiv{j\over n}\le 1,\quad
j=0,1,\dots n,\label{contin}
\end{equation}
and assume $\{\xi_j\}$ and $\{c_j\}$ are a discretization of some continuous functions, $\varphi(t)$
and $\psi(t)$ respectively, so that
\begin{equation}
\xi_j={\varphi(t)\over \sqrt n},\quad c_j={\psi(t)\over \sqrt n} 
\label{eq:contdef}
\end{equation}
[note in passing that $\varphi(t)\ge0$ and~$\psi(t)\ge0$ ].
%
%
%
%
The normalization condition for $\{\xi_j\}$ and~$\{c_j\}$ holds if we  impose 
\begin{equation}
\int_0^1 dt\,\varphi^2(t)=1,\quad \int_0^1 dt\,\psi^2(t)=1 .
\label{eq:norma}
\end{equation}
{}From~(\ref{eq:contdef}), we have $\xi_{j+1}-\xi_j\simeq n^{-3/2} [d\varphi(t)/dt]$, and~Eq.~(\ref{eq:action}) can be viewed as a discretized version of the functional $S[\varphi]$
, defined by
\begin{eqnarray}
&&\kern-1emS[\varphi]=\!{\varphi^2(0)+\varphi^2(1)\over 2n}\nonumber \\[.5em]
&&\kern-1em\phantom{S}+\!{1\over n^2}\!\!\int_0^1 \!\!dt\!\left[{1\over2 }\!\left(d\varphi\over dt\right)^2\!\!\!-{\omega^2\over2}\!\left(\varphi^2\!-\!1\right)\!+\!\sigma(\varphi-\lambda\psi)\right]\!,
\label{ebc13.06.12-1}
\end{eqnarray}
where $\omega$ is a positive constant (the properly scaled Lagrange multiplier: $\omega=n b$) and $\sigma(t)$ is a function that interpolates the set of multipliers $\{s_j\}$, i.e.,
\begin{equation}
s_j= n^{-5/2}\sigma(t).
\label{eq:contdef2bis}
\end{equation}
With this, Eq.~(\ref{eq:kktcond2}) becomes $\sigma(t)\ge 0$. Similarly,  the primal feasibility  conditions in~(\ref{sdp}) and the slackness condition~(\ref{eq:kktcond3}) become
\begin{eqnarray}
\varphi(t)-\lambda\psi(t)\phantom{]}&\le&0,\label{primal c}\\
\sigma(t)[\varphi(t)-\lambda\psi(t)]&=&0. \label{slackness c}
\end{eqnarray}
Note that by imposing the boundary conditions $\varphi(0)=0$ and $\varphi(1)=0$,
the functional $S[\varphi]$ becomes $O(n^{-2})$. 
%


More interestingly, the minimization of $S[\varphi]$ defines a mechanical problem, of which the second line in~Eq.~(\ref{ebc13.06.12-1}) is the  `action' and the corresponding integrant the `Lagrangian':   
\begin{equation}
L={1\over2 }\left(d\varphi\over dt\right)^2-{\omega^2\over2}\varphi^2+\sigma\varphi.
\end{equation}
It describes a driven harmonic oscillator with angular frequency $\omega$,
whose `equation of motion' is
\begin{equation}
{d^2\varphi\over dt^2}+\omega^2\varphi=\sigma .
\label{dho}
\end{equation}

To solve this problem, we first note that the
slackness conditions imply that either $\varphi(t)=\lambda\psi(t)$, in which case~$t$ is in the so called {\em coincidence set} $\mathscr C$, or  $\sigma(t)=0$. In the second  case, $t\in{\mathscr C}^c$ (${\mathscr C}^c$ stands for the complement of~$\mathscr C$), the primal feasibility condition is~$\varphi(t)<\lambda\psi(t)$, and Eq.~(\ref{dho}) becomes homogeneous (the equation of motion of a free harmonic oscillator). It has the familiar solution
\begin{equation}
\varphi(t)=A\sin\omega t+ B \cos\omega t ,\label{sho-no}
\end{equation}
where $A$, $B$ and $\omega$ are constants to be determined.
In the coincidence set~$\mathscr C$,~$\sigma$ is determined by~\eqref{dho}, where we make the substitution~$\varphi(t)=\lambda\psi(t)$ (recall that $\psi$ is a given function, as the components $c_j$ are themselves given). 
If we restrict ourselves to fiducial states $|\Psi_0\rangle$ whose  components $c_j$ are such that $\psi(t)$, defined through Eq.~(\ref{eq:contdef}), is continuous in the whole unit interval, one can show that the solution $\varphi(t)$ and its first derivative must be also continuous there [except in points of $\mathscr C$ where $\psi(t)$ itself is not differentiable]. Most of the physically relevant cases are of this type; some of them are considered in the examples below.
By taking into account the boundary conditions, as well as the continuity of $\varphi(t)$ and its derivative in the boundaries of $\mathscr C$, one can determine the arbitrary constants  that arise in solving the equation of motion. 
%
%
%

Before presenting  examples of this approach, we note that the minimum value of $S$ can be expressed in terms of the Lagrange multiplier (function) $\omega$ ($\sigma$), and the given function $\psi$, as
%
%
\begin{equation}
S_{\rm min}=
{1\over n^2}\left({\omega^2\over2}-{\lambda\over2}\int_0^1 dt\,
\sigma\psi\right)  .
\label{action}
\end{equation}
To prove this, we just have to integrate by parts~(\ref{ebc13.06.12-1}) and use the equation of motion~(\ref{dho}) and the boundary conditions $\varphi(0)=\varphi(1)=0$. Note that the integral is effectively over the coincidence set $\mathscr C$, where the expression for $\sigma(t)$ is given by: $\sigma=\lambda (d^2\psi/dt^2+\omega^2 \psi)$, as discussed above.


\subsection{\boldmath Large abstention ($\lambda\gg1$)}

For values of the abstention rate very close to one (large~$\lambda$), and provided $c_j> 0$ for all~$j$, the quantities~$\lambda c_j$ are also very large and~${\mathscr C}=\emptyset$. 
In this case $\sigma\equiv0$ in $[0,1]$, Eq.~(\ref{dho}) becomes homogeneous and we are dealing with a regular Sturm-Liouville eigenvalue problem. The solution is
\begin{equation}
\varphi(t)=A\sin\omega t;\quad  \omega=\pi m,\quad m=1,2,\dots ,\label{sho}
\end{equation}
where the boundary conditions $\varphi(0)=\varphi(1)=0$
have been taken into account to discard the independent $\cos\omega t$ solution.
Since we must have $\varphi(t)\ge0$ in the whole unit interval, we find that  $m=1$ (which gives the minimum eigenvalue of $d^2/dt^2$ for the given boundary conditions).
The constant $A$ is fixed by normalization and takes the value
$A=\sqrt2$, thus
\begin{equation}\label{solu 1}
\varphi(t)=\sqrt2\sin\pi t,
\end{equation}
namely $\xi_j\simeq \sqrt{2/n}\sin(\pi j/n)$. The minimum value of~$S$~is
\begin{equation}
S^*={\pi^2\over2n^2}.
\end{equation}
This leads to an asymptotic maximum fidelity of
\begin{equation}
F^{*}=1-{\pi^2\over4N^2} ,
\label{ebc29.04.12-3}
\end{equation}
which coincides with the known fidelity results for optimal phase encoding~\cite{BMM-T,fiurasek2}.

\subsection{\boldmath $|\Psi_0\rangle$ proportional to the POVM seed state $ |\Phi_1\rangle$}\label{phasePOVM}

The example we consider here is very simple from a computational point of view and yet illustrates that even a tiny rate of abstention can drastically improve the asymptotic fidelity $F$ of parameter estimation. More precisely, we will show that any finite amount of abstention enables changing the shot noise limit scaling $N^{-1}$ of~$1-F$ for large $N$  into the Heisenberg limit  scaling:~$N^{-2}$. 
 The elements of the family are equal superposition of all `Fock' states $\ket{j}$, i.e.
%
$ c_j=1/\sqrt{n+1}$. Despite of having such a large support, in the standard approach,~$Q=0$ ($\lambda=1$), the phase estimation fidelity these states provide does not exceed the shot noise limit: $1-F=1/(2N+2)$. This can be exactly computed for any $N$ with ease  from~\eqref{eq:phaseM}. Of~course it also agrees with the analytic asymptotic results: using Eq.~(\ref{eq:contdef}) we obtain $\varphi(t)=\psi(t)=1$, for $t\in[0,1]$, and the $1/n$ ($=1/N$)  boundary term in the action \eqref{action} is dominant.
%
%

Let us know address the more interesting case of $Q>0$ ($\lambda>1$). 
Here we can freely impose~$\varphi(0)=\varphi(1)=0$ and get rid of the shot-noise type term $1/n$.
In a sufficiently small neighbourhood  of $t=0$, i.e., for $0\le t<\alpha$, where~$\alpha $ is likewise small, we have $\varphi(t)-\lambda<0$, and the complementary slackness condition~(\ref{slackness c}) implies $\sigma(t)=0$ there. If $\alpha$ is the maximum value of $t$ less that $1/2$ for which this condition holds, it must be a boundary point of the coincidence set $\mathscr C$.  Then, for $t\geq\alpha$ the solution is given by the rescaled input state $\varphi(t)=\lambda\psi(t)=\lambda$. Thus,
%
\begin{equation}
\varphi(t)=\left\{
\begin{array}{ll}
A\sin \omega t ,&\quad 0\le t<\alpha;\\[.5em]
\lambda,&\quad \alpha<t\le 1/2,
\end{array}
\right.
\label{sol}
\end{equation}
where the constants $\alpha$, $\omega$ and $A$ are to be determined.
Continuity of $\varphi(t)$ and its derivative at $t=\alpha$ yields
\begin{equation}
A\sin\omega\alpha=\lambda,\qquad A\omega\cos\omega\alpha=0.
\end{equation}
We are left with the following possibilities for $\omega$ and $A$:
\begin{equation}
\omega\alpha\!=\!(2m\!+\!1){\pi\over2},\;\; A\!=\!(-1)^m\!\lambda;\; \;m=0,1,2,\dots .
\end{equation}
The positivity condition $\varphi(t)\ge0$ requires $m=0$, and normalization, Eq.~(\ref{eq:norma}), 
\begin{equation}
\alpha=1-{1\over\lambda^2}=Q.
\end{equation}
 Note that since $\alpha\le1/2$ we have $Q^{*}=1/2$. 
Combining these results we obtain
%
\begin{equation}
 \omega=\frac{\pi}{2Q}.\label{omega}
 \end{equation}
Extending the solution to the entire unit interval by applying the obvious symmetry of the problem, namely $\varphi(t)=\varphi(1-t)$, one has for $0<Q\le Q^*$ ($1<\lambda\le\sqrt2$)
\begin{equation}
\varphi(t)\!=\!\left\{\!\!
\begin{array}{lrcl}
\displaystyle {\bar Q^{-{1\over 2}}}\sin{\pi t\over2Q}, \; &0\!&\le t<&\! \displaystyle {Q};
\\[1em]
{\bar Q^{-{1\over 2}}}, &\displaystyle  {Q}&\!\le t\le\!& \displaystyle {\bar Q};
\\[.7em]
\displaystyle {\bar Q^{-{1\over 2}}}\sin{\pi (1\!-\!t)\over2Q}, \; &\displaystyle  {\bar Q}&\!< t\le\!& 1.
\end{array}
\right.
\end{equation}
%
Note that $\mathscr C=[Q,\bar Q]$ and $\sigma(t)=\omega^2\lambda$ for $t\in{\mathscr C}$ [$\sigma(t)=0$ for $t\in{\mathscr C}^c$]. Therefore, Eq.~(\ref{action}) gives
\begin{equation}
S_{\rm min}={\pi^2\over8 Q \bar Q \,n^2} ,\qquad 0<Q\le Q^*,\label{deltaCconst}\end{equation}
from which
\begin{equation}
F=1-{\pi^2\over16Q\bar QN^2} ,\qquad 0<Q\le Q^{*}=1/2.\label{deltaCconst2}
\end{equation}
For $1/2<Q \le1$ the solution is~(\ref{solu 1}) and the fidelity in~(\ref{ebc29.04.12-3}).
Note that even the slightest abstention rate unlocks the encoding power of the phase states 
and drastically changes the estimation precision from the original~$N^{-1}$  to $N^{-2}$.  

\begin{center}
\begin{figure}
	\centering
	\setlength{\unitlength}{5mm}
\thinlines
\begin{picture}(16,11)(0,0)
\put (1,1.2){\includegraphics[width=20em]{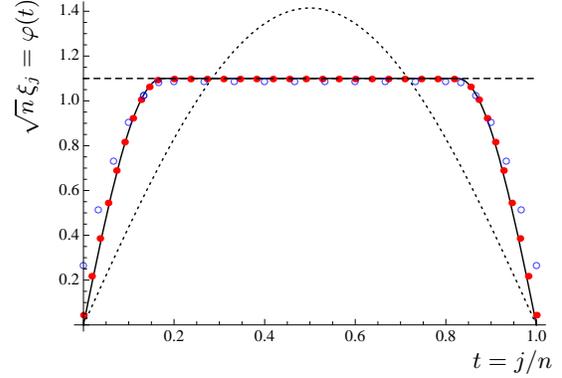}}
\put (-0.2,6.8){\rotatebox{90}{$\sqrt n\,\xi_j=\varphi(t)$}}
\put (12,.6){$t=j/n$}
\end{picture}
\caption{Profile of the components $\xi_j$ of the transformed fiducial state $|\tilde\Psi_0\rangle$  in the asymptotic limit. The solid line is the limiting function $\varphi (t)$ for $Q=0.17$ ($\lambda=1.1$). The points represent the actual components, as obtained by numerical optimization for $n=30$ (empty circles) and $n=220$ (filled circles). The dotted line is the solution for unrestricted abstention, Eq.~(\ref{solu 1}), whereas the dashed horizontal line represents the components $\psi(t)=1=\sqrt n c_j$ of the initial fiducial state $|\Psi_0\rangle$ scaled by $\lambda$.} \label{phivt} 
\end{figure}
\end{center}

The above results are illustrated in  Fig.~\ref{phivt}, where we represent the optimal solution for a 17\% abstention rate. 
Notice how the slackness conditions apply in the different regions: the straight part of~$\varphi$ (corresponding to $t\in{\mathscr C}$) is just $\lambda\psi=\lambda$, while the sinusoidal curves in the extremes (corresponding to the unconstrained region ${\mathscr C}^c$) smoothly match the straight line at the boundary.
The agreement between the numerical points and the analytic continuum limit is also quite evident.

\subsection{Multiple copies on the equator}\label{phases}
Let us now focus on phase estimation with a signal of the form
\begin{equation}
\ket{\Psi(g)}=\left(\frac{\ket{0}+\expo{i \theta}\ket{1}}{\sqrt{2}}\right)^{\!\!\otimes n} ,
\label{ebc14.06.13-1}
\end{equation} 
that is, with $N=n$ copies of states lying on the equator of the Bloch sphere. For these the coefficients $c_j$ read
\begin{equation}
c_j=2^{-\frac{n}{2}}\sqrt{\binom{n}{j}}.    \label{cks} 
\end{equation}
The maximum fidelity that can be attained with this signal without abstention is well known to be
$1-F=1/(4N)=1/(4n)$ for large $n$~\cite{BMM-T,fiurasek2}. To compute the effect of abstention  we proceed along the lines of the previous section.
In the asymptotic limit Eq.~(\ref{eq:contdef}) leads to
\begin{equation}
\psi(t)=\left[{n\over 2\pi t(1-t)}\right]^{1/4}\exp\left\{-\frac{n}{2}[\log 2-H(t)]\right\},\label{ccks}
\end{equation}
where $H(t)=-t\log t-(1-t)\log(1-t)$ is the Shannon entropy, and we have used Stirling's approximation.
Note that $\log 2-H(t)$ is the (binary) relative entropy~\mbox{$H(t\!\parallel \!1/2)$} between a Bernoulli distribution with success probability $p=t$ and the flat one~\mbox{($p=1/2$)}.
As in the previous case, the problem is invariant under~\mbox{$t\to1-t$}, which suggest using the variable~\mbox{$\tau=t-1/2$}, $\tau\in[-1/2,1/2]$, instead of $t$. Hence, the solution must be an even function of $\tau$.
In the region~$|\tau|\lesssim n^{-1/2}$ [i.e., around the peak of the distribution~(\ref{ccks})], we can use the Gaussian approximation
\begin{equation}
\psi(\tau)\approx\left(\frac{2n }{\pi}\right)^{\!\!1/4}\!\!\expo{-n\tau^2}, \label{gaussc}
\end{equation}
 where we slightly abuse notation here and in the rest of the section and use $\psi(\tau)$ to denote $\psi(t(\tau))$.  
 At the tails ($|\tau|>n^{-1/2}$), $\psi(\tau)$ falls off with an exponential rate given  by $H(1/2+\tau\!\parallel \!1/2)$. 
 
Since the solution of the minimization must be an even function of $\tau$, it must have the form
\begin{equation}
\varphi(\tau)=\left\{
\begin{array}{ll}
A\cos \omega\tau ,&\quad 0\le \abs{\tau}\le\alpha,\\[.5em]
\lambda\psi(\tau),&\quad \alpha<\abs{\tau}\le 1/2,
\end{array}
\right.
\label{sol2}
\end{equation}
%
%
The continuity of both $\varphi(\tau)$ and $\varphi'(\tau)$ at the boundary of ${\mathscr C}$, i.e., at the  point $\tau=\alpha$ read:
\begin{eqnarray}
A\cos(\Omega)&=&\lambda \psi
(\alpha),\label{contcond1}\\
-\Omega A\sin(\Omega)&=&\alpha\lambda \psi'(\alpha),\label{contcond2}
\end{eqnarray}
where we have defined $\Omega\equiv\omega\alpha$.
Combining these equations we obtain
\begin{eqnarray}
\Omega\tan\Omega&=&-\alpha{\psi'(\alpha)\over\psi(\alpha)},\label{cond1 man}\\
A^2&=&\lambda^2\left\{\psi^2(\alpha)+{\alpha^2\over\Omega^2}\left[\psi'(\alpha)\right]^2\right\}.
\label{cond2 man}
\end{eqnarray}
The normalization condition \eqref{eq:norma} turns out to be
%
%
\begin{equation}
A^2\frac{\alpha (2 \Omega+\sin2\Omega)}{2\Omega}+2\lambda^2\int_{\alpha}^{1/2}\!\!\!\psi^2(\tau) \,d\tau=1.\label{norm2}
\end{equation}

Eqs.~(\ref{contcond1}) through~(\ref{norm2})  cannot be solved analytically, but we can find asymptotic solutions by focusing on some specific regimes.  The first we will consider arises when the boundary points $\pm\,\alpha$ scale as $n^{-1/2}$, so that~${\mathscr C}$ stretches to the region around the peak of $\psi(\tau)$. In this case,~$\varphi(\tau)=\lambda\psi(\tau)$ gives the dominant contribution to~$S_{\rm min}$ and, as one intuitively expects, $S_{\rm min}\sim n^{-1}$. The two pieces of~$\varphi$ in Eq.~(\ref{sol2}) can be matched for arbitrary values of~$\lambda$ and the abstention rate can be finite (is not required to scale with $n$).
The second regime arises when $\alpha$ is fixed. In this situation, for sufficiently large $n$, the coincidence set~${\mathscr C}$ lies on the tails of $\psi(\tau)$. Matching the two pieces of $\varphi$ requires that $\lambda$ scales exponentially with~$n$, which means that the acceptance rate~$\bar Q$
must vanish also exponentially. In return, the piece of $\varphi$ in the first line of Eq.~(\ref{sol2}) has a wide (non vanishing) domain,~$[-\alpha,\alpha]$, and $S_{\rm min}\sim n^{-2}$ ($1-F\sim N^{-2}$), thus attaining the Heisenberg limit. Let us now consider the two regimes in more detail.
%
%
\subsubsection{$1/n$ regime}

%
\begin{figure}[b]
	\centering
	\setlength{\unitlength}{5mm}
\thinlines
\begin{picture}(16,11)(0,0)
\put (1,1.2){\includegraphics[scale=.75]{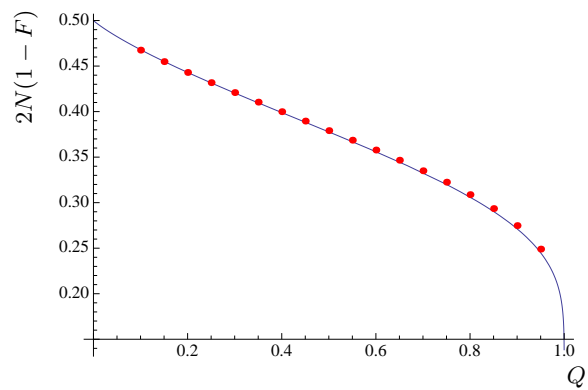}}
\put (-0.2,7.5){\rotatebox{90}{$2N(1-F)$}}
\put (14.5,.6){$Q$}
\end{picture}
\caption{Plot of $n S_{\rm min}=2N(1-F)$ vs $Q$ (solid line) for an asymptotically large number, $N=n$, of parallel spins on the equator of the Bloch sphere. The dots have been obtained by numerical optimization with $n=100$.}\label{SvsQ2} 
\end{figure}
%

We write $\alpha=a/\sqrt{n}$, where $a$ is fixed. 
Using the Gaussian approximation in Eq.~(\ref{gaussc}), Eqs.~\eqref{cond1 man} through~\eqref{norm2} become 
\begin{eqnarray}
a^2&=&\frac{\Omega\tan\Omega}{2},\label{as} \\[.5em]
A^2&=&\left({2n\over\pi}\right)^{1/2}\frac{\lambda^2\expo{-2a^2}(4a^4+\Omega^2)}{\Omega^2} , \label{A0} \\[.5em]
1&=&A^2\frac{a (2 \Omega+\sin2\Omega)}{2\sqrt n\,\Omega}\!+\!\lambda^2\left[1\!-\!{\rm Erf}(\sqrt{2}\,a)\right]\!,\label{norm3}
\end{eqnarray}
where ${\rm Erf}(x)$ is the error function.
Eq.~(\ref{norm3}) is correct up to exponentially vanishing contributions, which can be neglected here.
In deriving this equation  we also used that~${\rm Erf}(\sqrt{n/2})\to 1$ for large $n$. 
%
%
Substituting Eq.~(\ref{A0})  in Eq.~(\ref{norm3}) we obtain
\begin{equation}
{1\over \lambda^2}\!=\!{\rm Erfc}(\sqrt{2} a)\!+\!\frac{
 a  (4 a^4 \!+\! \Omega^2) (2 \Omega \!+ \!
    \sin2\Omega)}{\sqrt{2 \pi} \Omega^3}\expo{-2 a^2}\!,
    \label{1/lambda2}
 \end{equation} 
 where $\rm Erfc$ is the complementary error function, defined as ${\rm Erfc}(x)=1-{\rm Erf}(x)$.
 %
%
%
Finally, with the help of the Gaussian approximation~(\ref{gaussc}), we compute the minimum action from Eq.~(\ref{action}) and obtain
\begin{equation}\label{S equator}
 S_{\rm min}\!=\!{\omega^2\over2n^2}-{\lambda^2\over2n^2}\!\!\left[
(\omega^2\!\!-\!n){\rm Erfc}(\sqrt{2} a)\!+\!{4n^{\phantom{3\over2}}\kern-.6em a\over\sqrt{2\pi}}{\rm e}^{-2 a^2}
\!\right]\!\!.
\end{equation}
 Eqs.~(\ref{as}) and~(\ref{1/lambda2}), along with $\omega=\Omega\,\sqrt n/a$ and $Q=1-1/\lambda^2$, enable writing all variables in terms of the single parameter $\Omega$. By further substituting in Eq.~(\ref{S equator}) we obtain the curve $(Q,S_{\rm min})$ in parametric form:
 \begin{widetext}
 \begin{eqnarray}
  Q&=&{\rm Erf}\left(\sqrt{\Omega\tan\Omega}\,\right)-\left(\Omega\sec^2\Omega+\tan\Omega\right)\sqrt{{\tan\Omega\over\pi\,\Omega}}\,\expo{-\Omega\tan\Omega},\label{para1}\\  
  S_{\rm min}&=&{1\over2 n}\left[1 + \frac{\tan^2\Omega-\Omega\left(2\Omega-\tan\Omega\right)\sec^2\Omega}{2\,\Omega^2\sec^2\Omega+\sqrt{\pi\,\Omega\tan\Omega}\;{\rm Erfc}\left(\sqrt{\Omega\tan\Omega}\right)\,\expo{\Omega\tan\Omega}}\right]^{-1} .\label{para2}
 \end{eqnarray}
 \end{widetext}
Note that, as announced above, $1-F$ goes as $1/ N$.

\begin{figure}
	\centering
	\setlength{\unitlength}{5mm}
\thinlines
\begin{picture}(16,12)(0,0)
\put (1,1.2){\includegraphics[scale=.75]{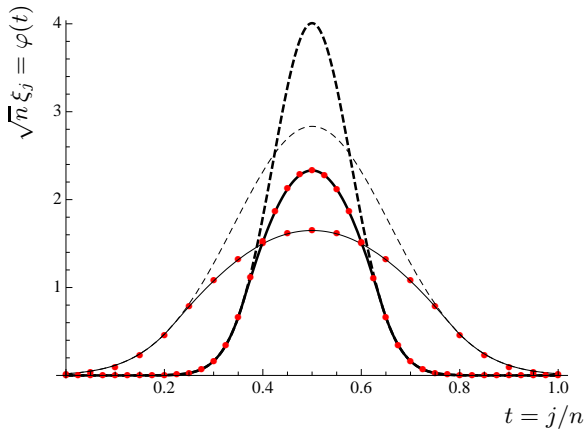}}
\put (-0.2,8.0){\rotatebox{90}{$\sqrt n\,\xi_j=\varphi(t)$}}
\put (13,.5){$t=j/n$}
\end{picture}
\caption{\label{prod} Profile of the transformed fiducial state $|\tilde\Psi_0\rangle$ 
for~\mbox{$Q=0.56$} \mbox{($\lambda=1.5$)}. The thin (thick) lines correspond to $n=20$ ($n=80$). The circles are obtained by numerical optimization. The dashed lines represent the constraint $\lambda\psi(t-1/2)$, where $\psi(\tau)$ is given in Eq,~\eqref{gaussc}.}
\end{figure}

 In Fig.~\ref{SvsQ2} we plot $n S_{\rm min}=2N(1-F)$ as a function of~$Q$, using Eqs.~(\ref{para1}) and~(\ref{para2}). The plot shows a strong dependence on $Q$. Hence, e.g.,  allowing about $90\%$ of abstention, has the same effect as doubling the number of copies in the standard approach (without abstention). Note also that for~$Q\to0$ we recover the well known result~\mbox{$2N(1-F)=1/2$}.
The profile of the transformed fiducial state $|\tilde\Psi_0\rangle$ is shown in Fig.~\ref{prod}, where $\varphi(\tau)$ and $\lambda\psi(\tau)$ are plotted as a function of $t=j/n$ for two different values of $n$ (recall that $\tau=t-1/2$).
%

\subsubsection{$1/n^2$ regime}

Here we assume that $\alpha$ is fixed (does not scale with~$n$). As $n$ goes to infinity, the boundaries of the coincidence set, $\tau=\pm\,\alpha$, lie on the tails of~$\psi(\tau)$, where the Gaussian approximation is not valid, and Eq.~(\ref{ccks}) must be used instead.
Eqs.~\eqref{cond1 man} and~\eqref{cond2 man} now become
%
 \begin{eqnarray}
 \Omega\tan \Omega &=&n\alpha\, {\rm arctanh}\,2\alpha-{2\alpha^2\over1-4\alpha^2}\label{wcon}, \\[.5em]
A^2\! &=&n^{{5/2}}\!\sqrt{\!{2\over\pi}}{\alpha^2\lambda^2{\rm arctanh}^22\alpha\over\Omega^2\sqrt{1\!-\!4\alpha^2}}\nonumber\\[.5em]
&\times&
\exp\!\left\{\!-n\!\left[\log2\!-\!H(\alpha\!+\!\mbox{\footnotesize${1\over2}$})\right]\right\} .
\label{wcon2}
 \end{eqnarray}
The first equation can be solved for $\Omega$ as an asymptotic series in powers of $1/n$:
\begin{equation}
\Omega={\pi\over2}+ O(n^{-1}), 
\end{equation}
which implies $\omega \simeq \pi/(2\alpha)$.
To evaluate the integral in Eq.~(\ref{norm2}), we expand the exponent $-n[\log2 -H(1/2+\tau)]$ around $\tau=\alpha$, so that
\begin{eqnarray}
&&\kern-.9em\int_{\alpha}^{{1\over2}}\!\!\psi^2(\tau)d\tau\approx\psi^2(\alpha)\int_{\alpha}^{{1\over2}}\!\!{\rm e}^{-2n(\tau-\alpha)\;{\rm arctanh}\,2\alpha}d\tau\nonumber
\\[.5em]
&&\kern-.9em\phantom{\int_{\alpha}^{{1\over2}}\!\!\psi^2(\tau)d\tau}\approx{1\over\sqrt{2\pi n}}{\exp\!\left\{\!-n\!\left[\log2\!-\!H(\alpha\!+\!\mbox{\footnotesize${1\over2}$})\right]\!\right\}\over \sqrt{1\!-\!4\alpha^2} \;{\rm arctanh}\,2\alpha}.
\label{offpeak}
\end{eqnarray}
We note that, although this contribution falls off exponentially exactly as $A^2$, it can be neglected in evaluating Eq.~(\ref{norm2}) since its prefactor is $O(n^{-1/2})$, as compared to that of $A^2$, which is $O(n^{5/2})$. Taking this into account and substituting $\Omega\approx\pi/2$ and Eq.~(\ref{wcon2}) into Eq.~(\ref{norm2}), we have
\begin{eqnarray}
A&=&{1\over\sqrt\alpha},\\[.2em]
\bar Q&=&{1\over\lambda^2}\approx
\left({2n\over\pi}\right)^{5/2}{\alpha^3\,{\rm arctanh}^22\alpha\over\sqrt{1-4\alpha^2}}\nonumber\\[.2em]
&&\phantom{{1\over\lambda^2}}
\times
\exp\!\left\{-n\!\left[\log 2\!-\!H(\alpha\!+\!\mbox{\footnotesize${1\over2}$})\right]\right\},
\label{falloff}
\end{eqnarray}
 and the critical acceptance rate is $\bar Q^*=2^{-n}$ (corresponding to $\alpha\to1/2$).
 
 The minimum action can be computed form Eq.~(\ref{action}) using the same approximation as in Eq.~(\ref{offpeak}). We obtain
 \begin{eqnarray}
 n^2 S_{\rm min}\!&=&\!{\pi^2\over8\alpha^2}\\[.5em]
 \!&-&\!{n^{{3\over2}}\over\!\!\sqrt{2\pi}}{\lambda^2{\rm arctanh}\kern.1em 2\alpha\over\sqrt{1-4\alpha^2}}
\! \exp\!\left\{\!-n\!\left[\log 2\!-\!\!H\!(\!\alpha\!+\!\mbox{\footnotesize${1\over2}$}\kern-.1em)\!\right]\!\right\}\!.\nonumber
 \end{eqnarray}
 Note that the exponential factor in the second line of this equation is cancelled by~$\lambda^2$, given in Eq.~(\ref{falloff}), and only the product of the pre-factors, of order $n^{-1}$, remains. Thus the second line can be safely neglected in the asymptotic limit and we have
%
 \begin{equation}
 F=1-\frac{\pi^2}{16N^2\alpha^2}+O(N^{-3}),\quad 0<\alpha\le1/2,
 \label{ebc06.06.13-1}
 \end{equation}
 with an abstention rate given by Eq.~(\ref{falloff}).
The maximum fidelity is attained by the largest value of~$\alpha=1/2$, for which $F=F^*$ as it should be. 
In summary, 
high abstention rate (exponentially small acceptance rate)  enables a drastic change in the scaling with the number of copies of the estimation precision. With such rates, one can attain $1-F\sim 1/N^2$, i.e., achieve the Heisenberg limit.

\section{Direction estimation}


%

Proceeding along the same lines as in Sec.~\ref{Sec-AsRePhEs}, we can write $\langle\xi|{\mathsf M}|\xi\rangle$ [recall Eq.~(\ref{fidelity-general-3})] as
%
\begin{equation}
\langle\xi|{\mathsf M}|\xi\rangle=\sum_{j=1}^{n}\frac{2j}{\sqrt{4j^2-1}}\xi_j\xi_{j-1},\label{dirmat}
\end{equation}
%
and $S=1-\langle\xi|{\mathsf M}|\xi\rangle$ becomes now
\begin{equation}
S\!=\!
{1\over2}\!\!\left[\sum_{j=1}^{n}j\!\!\left({\xi_{j}\over \sqrt{j\!+\!{1\over2}}}-{\xi_{j-1}\over\sqrt{ j\!-\!{1\over2}}}\right)^2+{(n+1)\,\xi_{n}^2\over n+{1\over2}}\right] ,
\end{equation}
where we have used the normalisation constraint in Eq.~(\ref{norm}). Introducing Lagrange multipliers according to KKT, and assuming $N=2n$ asymptotically large, we obtain the equivalent variational problem of minimizing the action
%
%
%
\begin{eqnarray}\label{diraction}
&&\kern-1.7emS\!=\!{\varphi^2(1)\over2n}\\
&&\kern-1.7em\phantom{S}+\!{1\over n^2}\!\!\int_0^1\!\!\!dt\! \left\{\!{t\over2}\!\left[{d\over d t}\!\!\left(\!\varphi\over\sqrt t\!\right)\right]^2\!\!\!\!-\!{\omega^2\over2}\!\left(\varphi^2\!\!-\!1\right)\!+\!\sigma(\varphi\!-\!\lambda\psi)\!\!\right\}\!,
\nonumber
\end{eqnarray}
%
where the primal feasibility condition~(\ref{primal c}) and the slackness condition~(\ref{slackness c}) still apply. 
For $\lambda=1$ no transformation of the state is possible, therefore the first,  order $n^{-1}$,  term in \eqref{diraction}  is fixed by the boundary value of the initial state $\psi(1)$. For $\lambda>1$ we can
impose $\varphi(1)=0$, hence opening the door to order~$n^{-2}$ scaling (i.e., to attaining the Heisenberg limit). 

The evolution equation corresponding to the second line in Eq.~(\ref{diraction}) is more conveniently expressed in terms of
$\tilde\varphi(t)=\varphi(t)/\sqrt t$. It reads
%
\begin{equation}
t^2{d^2\tilde\varphi\over dt^2}+t{d\tilde\varphi\over dt}+\omega^2 t^2\,\tilde\varphi=t^{3/2}\;\sigma .
\label{difeqdir}
\end{equation}
The minimum value of the action can be written as in Eq.~(\ref{action}),
%
%
where we recall that $\sigma(t)$ can be only different from zero in the coincidence set $\mathscr C$. Now, $\sigma(t)$ is given by Eq.~(\ref{difeqdir}) with $\tilde\varphi(t)=\lambda\psi(t)/\sqrt t$.

\subsection{\boldmath Large abstention ($\lambda\gg1$)}

For abstention rates close to unity, and provided $c_j>0$ for all $j$, one has ${\mathscr C}=\emptyset$, so $\sigma(t)\equiv0$. Eq.~(\ref{difeqdir}) becomes homogeneous and its solution is
\begin{equation}
\varphi(t)=A \sqrt t\,J_0(\omega t)+ B\sqrt t\,Y_0(\omega t),
\end{equation}
where $J_0$ and $Y_0$ are Bessel functions of first and second kind respectively, and $A$, $B$ and $\omega$ are constants that we fix by requiring $\varphi(1)=0$ (otherwise $S$ is order $1/n$) and the convergence of the integral in Eq.~(\ref{diraction}). The latter implies $B=0$. 
%
The former condition and the positivity of $\varphi(t)$ fixes $\omega$ to be the first zero of~$J_0$, which we call~$\gamma_{1}$. 
Hence,~$\omega=\gamma_1\approx  2.405$.
%
%
Imposing normalization  we finally fix $A$, and the solution is
\begin{equation}
\varphi(t)={\sqrt{2t}\over J_1(\gamma_1)}J_0(\gamma_1 t) .
\label{eq:optdirphi}
\end{equation}
%
%
Using Eq.~\eqref{action}, we obtain $S^{*}={\gamma_1^2/ 2 n^2}$, and the maximum fidelity is
\begin{equation}
F^{*}=1-{\gamma_1^2\over N^2} ,
\label{eq:diropsol}
\end{equation}
in agreement with~\cite{bagan-optimal}. The abstention rate required to achieve the Heisenberg limit strongly depends on the initial family of states, as will be shown in the following two examples.

\subsection{\boldmath $|\Psi_0\rangle$ proportional to the POVM seed state $ |\Phi_2\rangle$}\label{directionPOVM}
\begin{center}
\begin{figure}
	\centering
	\setlength{\unitlength}{5mm}
\thinlines
\begin{picture}(16,12)(0,0)
\put (1,1.2){\includegraphics[scale=.75]{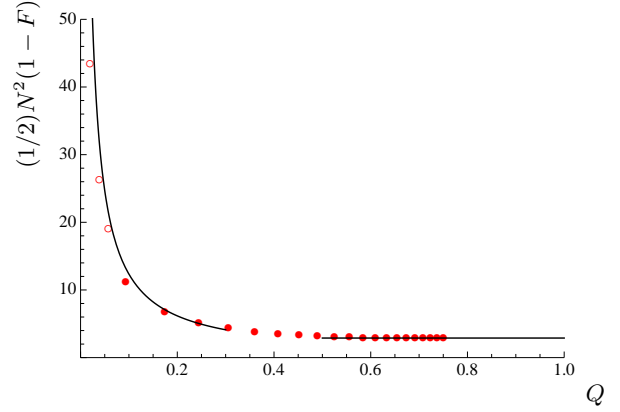}}
\put (-0.2,6.8){\rotatebox{90}{$(1/2) N^2(1-F)$}}
\put (15,.6){$Q$}
\end{picture}
\caption{Plot of $n^2S_{\rm min}[=(1/2)N^2(1-F)]$ versus $Q$. The solid lines are  the analytical expressions in~(\ref{SdirPOVM}), whereas the circles are numerical results. In order to approach the asymptotic limit, higher values of $n$ are needed for smaller $Q$. Accordingly, two different values of $n$ have been used; $n=50$ (filled circles) and $n=120$ (empty circles).
}\label{dirPOVM} 
\end{figure}
\end{center}

In analogy with Sec.~\ref{phasePOVM}, in this example we choose the fiducial state $\ket{\Psi_0}$ to be proportional to the POVM seed $\ket{\Phi_2}$ in Eq.~\eqref{seeds}. This leads to~$\psi(t)=\sqrt{2t}$, and the solution has the form
\begin{equation}
\varphi(t)\!=\!
\left\{\kern-.3em
\begin{array}{ll}
\lambda\sqrt{2t}, &\; 0\le t\le\alpha, \\[.5em]
A\,\sqrt{t}\,J_0(\omega \,t)\!+\!B\,\sqrt{t}\,Y_0(\omega\, t), &\;\alpha<t\le1.
\end{array}
\right.
\end{equation} 
Then, $\sigma(t)=\lambda\omega^2\sqrt{2t}$, if $t\in{\mathscr C}=[0,\alpha]$ (and it vanishes otherwise).
Substituting in Eq.~(\ref{action}), the minimum action can be written as
\begin{equation}\label{SdirPOVM-1}
S_{\rm min}={\omega^2\over2 n^2}(1-\alpha^2\lambda^2).
\end{equation}
Continuity of $\varphi(t)$ and its first derivative at $t=\alpha$, imply
\begin{equation}
A=-\frac{\pi\alpha\lambda\omega}{\sqrt{2}}Y_1(\omega\alpha),\quad
B=\frac{\pi\alpha\lambda\omega}{\sqrt{2}}J_1(\omega\alpha) ,
\end{equation} 
and the boundary condition $\varphi(1)=0$ requires, 
 \begin{equation}
J_1(\omega\alpha)Y_0(\omega)-Y_1(\omega\alpha)J_0(\omega)=0\label{bcon} .
\end{equation}
We will not attempt to find the exact analytical solution of this transcendental equation, but rather, consider two particular regions of $\alpha$ (the boundary of the coincidence set $\mathscr C$) where approximate solutions can be easily derived. They are given by \mbox{$\alpha\gtrsim 0$} and 
$\alpha\lesssim 1$. That will suffice to capture the main features of $S_{\rm min}$ (see Figure \ref{dirPOVM}). 
Note that small $\alpha$ corresponds to large $\lambda$, since the coincidence set ${\mathscr C}=[0,\alpha]$ is a small region and thus $\varphi(t)$ cannot differ much from the unconstrained solution that leads to $F^*$. On the other hand, 
$\alpha\lesssim 1$ must correspond to small abstention.

%
%
%
If $\alpha\gtrsim 0$, we substitute the ansatz $\omega=\gamma_1+a\alpha+b\alpha^2+\ldots$ in~(\ref{bcon}). After some algebra, we obtain
%
\begin{equation}
\omega=\gamma_1\left[1+\frac{\alpha^2}{2J^2_1(\gamma_1)}+O(\alpha^4\log{\alpha})\right],\label{wasmall}
\end{equation}
where we have made use of the relation
\begin{equation}\label{rel JY}
J_1(z)Y_0(z)-Y_1(z)J_0(z)={2\over\pi z},\quad \mbox{for all $z$};
\end{equation}
in particular,~$Y_0(\gamma_1)=2J^{-1}_1\!(\gamma_1)/(\pi\gamma_1)$.

If $\alpha\lesssim 1$, Eq.~(\ref{bcon}) can only hold for very large $\omega$ and~$\alpha\omega\approx\omega$, 
as is apparent from Eq.~(\ref{rel JY}), and we can replace the Bessel functions for their well known asymptotic approximations
\begin{eqnarray}
J_k(z)&\approx&\sqrt{\frac{2}{\pi z}}\cos\left(z-\frac{k\pi}{2}-\frac{\pi}{4}\right),\nonumber\\
Y_k(z)&\approx&\sqrt{\frac{2}{\pi z}}\sin\left(z-\frac{k\pi}{2}-\frac{\pi}{4}\right).
\end{eqnarray}
 With this, Eq. \eqref{bcon} becomes
\begin{equation}
\frac{2\cos{\omega(1-\alpha)}}{\pi\omega\sqrt{\alpha}}=0,
\end{equation} 
from which
\begin{equation}
\omega=\frac{\pi}{2(1-\alpha)}.
\end{equation} 
We next impose the normalisation condition to find the relationship between $\lambda$ and $\alpha$. For $\alpha\gtrsim0$, we find
\begin{equation}\label{islambda}
\lambda^2={1\over J^2_1(\gamma_1)}+O(\alpha^2\log\alpha).
\end{equation}
Taking the limit $\alpha\to0$ we find  the critical value of~$\lambda$: \mbox{$\lambda^*=1/J_1(\gamma_1)$}; and the critical rate of abstention:
\begin{equation}
Q^*=1-J_1^2(\gamma_1)\approx0.73.
\end{equation}
Substituting Eq.~(\ref{wasmall}) and~(\ref{islambda}) in Eq.~(\ref{SdirPOVM-1}) we readily see that the various contributions to order $\alpha^2$ cancel, and
\begin{equation}
S_{\rm min}={1\over n^2}\left[{\gamma^2_1\over2}+O(\alpha^4\log\alpha)\right].
\end{equation}
One can check that, as expected,  $S_{\rm min}$ (and thus the fidelity) is flat in the region $\alpha\gtrsim0$ ($Q\lesssim Q^*$); i.e.,~$S_{\rm min}$ is a smooth function of $Q$ at $Q=Q^*$. Indeed, Eq.~(\ref{islambda}) implies \mbox{$\alpha^2=o(\lambda^{*2}-\lambda^2)=o(Q^*-Q)$}, and~\mbox{$n^2S_{\rm min}=\gamma^2_1/2+o[(Q^*\!\!-\!Q)^2\!\log(Q^*\!\!-\!Q)]$}. The correction can be computed explicitly with some effort. We find that $S_{\rm min}$ increases up to $3.5\%$ for $Q\approx 0.6$, at which point the approximation breaks down.

For $\alpha\lesssim1$, we find
\begin{equation}
\lambda^2\approx{1\over\alpha}.
\end{equation}
Combining all these results, we find
%
%
\begin{equation}
n^2S_{\rm min}\approx\left\{
\begin{array}{ll}\displaystyle
\frac{\pi^2}{8Q},&\; Q\gtrsim0\\[1em]
\displaystyle
\frac{\gamma_1^2}{2},&\; Q\lesssim Q^* ,
\end{array}
\right.\label{SdirPOVM}
\end{equation}
where we insist that this expression is a very good approximation down to relatively small values of $Q$, as can be seen in Fig.\ref{dirPOVM}. 
In this figure we plot~Eq.~\eqref{SdirPOVM} for each regime (lines), along with some numerical results (points). The plot shows a very good agreement for most of the values of the abstention rate~$Q$. 
One can see that the flat region extends to values of $Q$ fairly smaller than $Q^*$.
Note again, that any nonzero amount of abstention enables the estimation accuracy to change behaviour from $1/N$ to $1/N^2$, thus attaining the Heisenberg limit. 

\subsection{Antiparalel spins}

As for the case of phase estimation, here we focus on signals consisting in product states of $N=2n$ spins. The simplest possibility is, of course, identical copies. However, this case is of no relevance to direction estimation with abstention, since the seed state $|\Psi_0\rangle$ has only a single component in the symmetric subspace of~$j=n$, i.e.,~$c_j=0,$ if $0\le j<n$, and abstention can only change the components by a multiplicative factor, as shown in Eq.~(\ref{transformed}). Thus $\xi_j=0,$ if $0\le j<n$ and $\xi_n= c_n$. 
%
%
Instead, we consider a seed state consisting of $2n$ antiparallel spins; $n$ of them pointing along the positive $z$-axis and the other $n$ pointing along the opposite direction,
\begin{equation}
\ket{\Psi_0}=\ket{\overbrace{\uparrow\uparrow\ldots\uparrow}^{n}\overbrace{\downarrow\downarrow\ldots\downarrow}^{n}}=\sum_{j=0}^{n}c_j\ket{j,0} .
\end{equation}
Such state has zero magnetic number, $m=0$ and non-vanishing components $c_j$ given by
\begin{equation}
c_j=\braket{\mbox{\footnotesize${n\over2}$},\mbox{\footnotesize${n\over2}$};\mbox{\footnotesize${n\over2}$},-\mbox{\footnotesize${n\over2}$}}{j,0}=n!\sqrt{\frac{2 j+1}{(n-j)!(n+j+1)!}}, 
\end{equation}
where 
$\braket{j,m;j',m'}{J,M}$ are the standard Clebsch-Gordan coefficients.
%
The `continuous version' of these components is given by ($t=j/n$)
\begin{equation}
\psi(t)\!=\!\sqrt{\frac{2 n t}{\!\!\!(1\!+\!t)\sqrt{1\!-\!t^2}}}\exp\!\left\{\!-n\!\!\left[\log{\!2}\!-\!H\!\!\left(\!\frac{1\!-\!t}{2}\!\right)\!\right]\!\right\}\!,
\label{phi}
\end{equation}
which has a peak at $t=0$.
The solution to the minimisation problem in Eq.~(\ref{diraction}) has the form
\begin{equation}
\varphi(t)=\left\{
\begin{array}{ll}
A\sqrt t\,J_0(\omega t),&\quad 0\le t\le\alpha;\\[.5em]
\lambda\,\psi(t),&\quad \alpha<t\le 1.
\end{array}
\right.
\label{soldir}
\end{equation}
Following the same lines as in Sec.~\ref{phases}, we consider two scalings of the boundary point $t=\alpha$: one where it goes to zero as $1/\sqrt n$, and a second one, where $\alpha$ is fixed. These will lead to two regimes, where $1-F$ vanishes respectively as $N^{-1}$ and $N^{-2}$.
%
\subsubsection{1/n regime}

\begin{center}
\begin{figure}
	\centering
	\setlength{\unitlength}{5mm}
\thinlines
\begin{picture}(16,12)(0,0)
\put (1,1.2){\includegraphics[scale=.75]{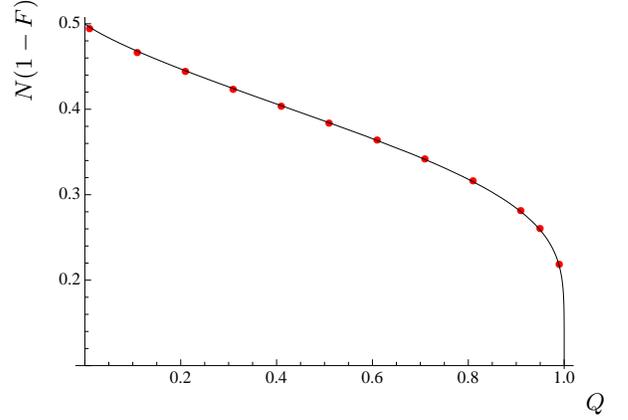}}
\put (-0.2,8.8){\rotatebox{90}{$N(1-F)$}}
\put (15,.6){$Q$}
\end{picture}
\caption{Plot of $nS_{\rm min}$[$=N(1-F)$] vs. $Q$ (solid line) for a signal state consisting of an asymptotically large number, $N=2n$, of antiparallel spins with null total magnetic number. The dots have been obtained by numerical optimization with~$n=100$.}\label{nSvsQ} 
\end{figure}
\end{center}

In this regime we set $\alpha=a/\sqrt n$. As in the phase case, we can use the `Gaussian approximation' for~(\ref{phi}):
\begin{equation}
\psi(t)=\sqrt{2\, n\, t}\,\expo{-n\, t^2/2}.\label{psiap}
\end{equation}
Note that
%
$
\int_0^1\psi^2(t)\,dt=1
$,
%
up to contributions that vanish exponentially with $n$.
The following expressions follow from the conditions of continuity of the solution and its derivative as well as normalisation:
%
%
%
%
%
\begin{eqnarray}
a^2&=&\Omega \;\frac{J_1(\Omega)}{J_0(\Omega)},\label{abeq}\\
A&=&\frac{\sqrt{2 n}\,\lambda\,\expo{-a^2/2}}{J_0(\Omega)},\\
\bar Q&=&{1\over \lambda^2}=\left(1+a^2+{a^6\over\Omega^2}\right){\rm e}^{-a^2},
\label{Qbar dir}
\end{eqnarray}
%
%
where we have defined $\Omega\equiv \omega\alpha=\omega a/\sqrt n$.  The minimum action  $S_{\rm min}$ is given by
%
%
%
%
%
\begin{equation}
S_{\rm min}=\frac{\Omega^2}{2n}\;\frac{1-a^2+a^4+\Omega^2}{a^6+(1+a^2)\Omega^2} ,
\end{equation}
where we have neglected exponentially vanishing terms. This expression, together with Eqs.~\eqref{abeq} and~(\ref{Qbar dir}) defines the curve $(Q,S_{\rm min})$ in terms of the free parameter~\mbox{$\Omega\in[0,\gamma_1)$}. The corresponding plot is shown in Fig.~\ref{nSvsQ}.
We see that for moderate values of the abstention rate one can substantially improve the estimation precision. E.g.,  a rate of abstention of $95\%$ has the same effect as doubling the number of spins in the standard approach (without abstention). Note, however, that with finite acceptance rate we cannot beat the shot noise limit.

\subsubsection{$1/n^2$ regime}
Here we take $\alpha$ to be fixed. From Eq.~\eqref{soldir} , continuity of $\varphi(t)$ and~$\varphi'(t)$ at $t=\alpha$ yield 
\begin{eqnarray}
A \sqrt{\alpha}\,J_0(\Omega)&=&\lambda\;\psi(\alpha),\label{contdir}\\
{A\over2\sqrt\alpha} \left[ J_0(\Omega)-2\Omega J_1(\Omega)\right] &=&\lambda\;\psi'(\alpha),\label{contderdir}
\end{eqnarray}
where, as before, $\Omega=\alpha\omega$.
It follows that
\begin{equation}
\frac{J_0(\Omega)\!-\!2\Omega J_1\!(\Omega)}{2\alpha\,J_0(\Omega)}\!=\!\frac{\psi'(\!\alpha)}{\psi(\alpha)}
\!=\!-n\;{\rm arctanh}\,\alpha\! +\! O(n^0), \label{aux}
\end{equation}
where Eq.~\eqref{phi} has been used. We can solve this equation for $\Omega$ as a series in inverse powers of $n$, obtaining 
\begin{equation}
\Omega=\gamma_1+O(n^{-1}),
\end{equation}
where we recall that $\gamma_1$ stands for the first zero of the function $J_0(z)$.
Substituting this result into Eq.~(\ref{contderdir}) we obtain
\begin{eqnarray}
A&=&{\sqrt{2}\,n^{3\over2}\alpha\lambda\;{\rm arctanh}\,\alpha\over (1-\alpha)^{1\over4}(1+\alpha)^{3\over4}\gamma_1 J_1(\gamma_1)}
\\[.5em]
&\times&\exp\left\{-n\left[\log2-H\left(\frac{1-\alpha}{2}\right)\right]\right\}.
\end{eqnarray}
Neglecting the contribution from the coincidence set, by the same arguments as in the paragraph after Eq.~(\ref{offpeak}),
the normalization condition is 
\begin{equation}
A={\sqrt2\over\alpha J_1(\gamma_1)} .
\end{equation}
%
%
%
Combining the last two equations, we find
%
 \begin{equation}
 \bar{Q}={1\over\lambda^2}\sim n^3\exp\left\{-2 n\left[\log 2-H\left(\frac{1-\alpha}{2}\right)\right]\right\}.
 \end{equation}
As for phase estimation, the acceptance rate $\bar Q$ falls off exponentially.
The minimum action $S_{\rm min}$ can be computed  from Eq.~\eqref{action} along the same lines as in the analogous phase estimation example. This leads to%
%
%
\begin{equation}
F=1-\frac{\gamma_1^2}{N^2\alpha^2}.
\end{equation}
As in Sec.~\ref{phases}, abstention enables exceeding the shot noise limit. Note that for $\alpha=1$,
we have \mbox{$F=F^*$}, Eq.~\eqref{eq:diropsol}, as expected.

\section{Frame estimation}

As anticipated in the introduction, if the encoding system consists of $N$ qubits one can make use of the multiplicities of the different irreducible representations (i.e. the degeneracy of the $j$ quantum number) to provide a very efficient encoding of the orientation of a Cartesian frame, or equivalently, of the rotation group parameters~$g$. States of the form 
$|\Psi_0\rangle =\sum_{j=0}^{n} c_j \,(\sum_{m=-j}^j|j,m,\alpha_m\rangle)/\sqrt{2j+1}$ 
exploit optimally these ancillary degrees of freedom and lead to a matrix~$\mathsf M$ that is (almost) equal to that corresponding to phase estimation. Hence, most of the expressions and conclusions derived in Section \ref{Sec-AsRePhEs} also hold in this case, but one must recall that $N=2n$ for frames (whereas $N=n$ for phases), i.e.  one must perform the change~$N\to N/2$ in the formulae of that section to obtain the corresponding formulae for frames. In particular, Eq.~(\ref{ebc29.04.12-3}) becomes~$F^*=1-\pi^2/N^2$ for frame estimation, in agreement with~\cite{framesopt};  Eq.~(\ref{ebc06.06.13-1}) becomes $F=1-\pi^2/(4N^2\alpha^2)+\dots$, and so on.
Note in particular that direction estimation does not provide an optimal strategy for frame estimation, namely,  the optimal frame fidelity cannot be attained by splitting the~$N$ qubits in three groups, encoding each orthogonal direction in one of them, and performing three independent direction estimations.

In our final example 
we move away from the $N$-qubit encoding towards a scenario where the degeneracy of the angular momentum representations cannot be used to  improve the frame estimation accuracy, as is the case of, e.g., an atom in a Rydberg state. 
In this scenario, we have
\begin{eqnarray}
\langle\xi|{\mathsf M}|\xi\rangle\!&=&\!1\!-\!{1\over2}\!\sum_{j=1}^n\!\left({1\over j\!+\!1}\!+\!{1\over j\!+\!{1\over2}}\right)\!\xi_j^2\!-\!\xi_0^2\!-\!{\xi_n^2\over2}\\
&-&\!{1\over2}\sum_{j=0}^{n-1}\!\left(\!j\!+\!{1\over2}\right)\!\!\left({\xi_{j+1}\over \sqrt{j\!+\!1\!+\!{1\over 2}}}\!-\!{\xi_{j}\over \sqrt{j\!+\!{1\over 2}}}\right)^{\!\!2}.\nonumber
\end{eqnarray}
%
In the asymptotic limit, the continuous version of this expression is cast as $\langle\xi|{\mathsf M}|\xi\rangle=1-S$, with the action
\begin{eqnarray}
S&=& {1\over n^2}\int^{1}_{0}dt\left\{ \frac{t}{2}\left[\frac{d}{dt}\left(\frac{\varphi}{\sqrt{t}}\right)\right]^2\right.\nonumber\\&+&\left.2n\frac{\varphi^2}{2t}-\frac{\omega^2}{2}\left(\varphi^2-1\right)+\sigma(\varphi-\lambda\psi)\right\},
\label{framesS}
\end{eqnarray}
which includes the constraints~(\ref{norm}) and~(\ref{sdp}) and the corresponding Lagrange multipliers $\omega$ and $\sigma$, and where we have set $\varphi(0)=\varphi(1)=0$. This action and that for direction estimation, Eq.~(\ref{diraction}), look much the same but for the term proportional to $n$. This apparently minor difference leads however to very different asymptotic behaviors.
%
%
The equation of motion that follows from~(\ref{framesS}) turns out to be
\begin{equation}
{d^2\over d t^2}\varphi+\left(\omega^2-\frac{2 n}{t}+{1\over 4t^2}\right)\varphi=\sigma.\label{edframes}
\end{equation}
Since $n$ is assumed to be asymptotically large, the term proportional to $n$ in~(\ref{framesS}) forces $\varphi(t)$ to peak at~\mbox{$t\approx1$} in order to minimize the action. Therefore, the last term in~(\ref{edframes}) can be safely neglected. 
The minimum value of~$S$ can be written in terms of the Lagrange multipliers and $\psi(t)$ as in Eq.~(\ref{action}), with $\sigma=\lambda[\psi''+(\omega^2\psi-2n/t)\psi$] for $t\in{\mathscr C}$, and $\sigma=0$ otherwise.

\subsection{Large abstention}

Once again, for abstention rates close to one, and provided $c_j\not=0$ for all $j$, Eq.~(\ref{edframes}) becomes homogeneous, i.e., $\sigma=0$, and, along with the boundary conditions~$\varphi(0)=\varphi(1)=0$, defines an eigenvalue problem. Its solution can be given in terms of Whittaker functions, but unfortunately is rather involved. It proved much simpler to formulate and solve a less demanding eigenvalue problem with the same large $n$ asymptotic behavior, as we explain next.

Since $\varphi(t)$ is peaked at $t\approx1$, we can Taylor expand the term $2n/t$ in Eq.~(\ref{edframes}) around this point. 
The leading and sub-leading contributions to $S_{\rm min}$ come from the first two terms in this expansion. That is, from the linear approximation: $2n/t\approx 2n+2n(1-t)$. Within this approximation the equation of motion becomes
\begin{equation}
{d^2\over d t^2}\varphi+2nt\varphi+(\omega^2-4n)\varphi=0,
\qquad
\varphi(1)=0,\label{difairi}
\end{equation}
and we relax the boundary condition $\varphi(0)=0$ by requiring only $\varphi(t)$ to vanish as $t\to-\infty$. This may seem unnatural at first, but it will become immediately apparent that the solution to this well-posed Sturm-Liouville eigenvalue problem vanishes exponentially with $n$ if $t\le0$ (in particular $\varphi(0)\to0$ exponentially as $n\to\infty$), which is enough to ensure that the resulting asymptotic expansion of  $S_{\rm min}$ in inverse powers of $n$ will be correct. 
Such solution~is:
\begin{equation}
\varphi(t)=C\; {\rm Ai}\left[{4n-\omega^2-2n t\over (2n)^{2/3}}\right],
\label{airy}
\end{equation}
where $\rm Ai$ is the Airy function and the constant $C$ is fixed by normalization
.
Imposing the second boundary condition, $\varphi(1)=0$, we have (for the smallest eigenvalue) $\omega^2=2n-\gamma_1\left(2n\right)^{2/3}$, where in this section $\gamma_1$ stands for the first zero of  ${\rm Ai}(x)$, whose value is $\gamma_1\approx -2.33811$. Using~(\ref{action}), we obtain the minimum action
\begin{equation}
S^*={1\over n}-{\gamma_1\over2^{1/3}n^{4/3}}+O(n^{-5/2}),
\label{ebc07.06.13-1}
\end{equation}
from which (recall that here $N=2n$)
\begin{equation}
F^*=1-{1\over N}+{\gamma_1\over N^{4/3}}+O(N^{-5/3}).\label{T}
\end{equation}
For the average of the error $e_3$ with which we estimate the three axes of the Cartesian frame (see Introduction), we obtain
$\langle e_3\rangle=8/N-8\gamma_1/N^{4/3}+O(N^{-5/3})$.
These results are in complete agreement with those in~\cite{frames}.

The asymptotic series we have obtained turns out to be in powers of $N^{-1/3}$. To obtain accurate values~of~$F^*$ for moderately large $N$, the next term in~(\ref{T}), of order $N^{-5/3}$,  might be important. Using our approach the calculation of this term is straightforward. One simply needs to include in~(\ref{difairi}) the next term in the Taylor expansion of $2n/t$, i.e., $2n(1-t)^2$, and use perturbation theory to obtain the correction \mbox{$\delta \omega^2=2n\int_{-\infty}^1 (1-t)^2\varphi^2(t)\, dt$}. The corresponding correction to~$S^*$ can then be computed via Eq.~(\ref{action}). The result is~$\delta S^*=2^{7/3}\gamma_1^2/(15 n^{5/3})$. From this, the correction to the fidelity turns out to be $\delta F^*=8\gamma_1^2/(15 N^{5/3})$.

\subsection{Limited Abstention}
%
As in the previous examples, if the rate of abstention is fixed to a value strictly less than one the resulting precision very much depends on the given signal state, namely, on the shape of $c_j$ (or $\psi$). In order to give a concrete expression for the fidelity, here we will assume that, maybe because of some energy limitations, the probability amplitudes $c_j$ of exciting a state (e.g., of a Rydberg atom) with angular momentum $j$  is a decreasing function. Let us further assume as a first approximation, and also for simplicity, that  this decrease is linear: $c_j\propto n-j$, which implies $\psi(t)=\sqrt3(1-t)$. This~simple example will allow us to illustrate the most characteristic features of frame estimation enhanced by abstention.

If no abstention is allowed (standard estimation), one can show that the averaged error [i.e., $8(1-F)$] vanishes as $ (1/N)\log N$ as $N$ increases, much slower than using the optimal signal states. We will show that even a tiny amount of abstention is enough to turn this scaling into~$1/N$. Moreover, the coefficient in this scaling law can be reduced down to almost the minimum value in~(\ref{T}) with a finite amount of abstention.

\begin{center}
\begin{figure}
	\centering
	\setlength{\unitlength}{5mm}
\thinlines
\begin{picture}(16,12)(0,0)
\put (1,1.2){\includegraphics[scale=.75]{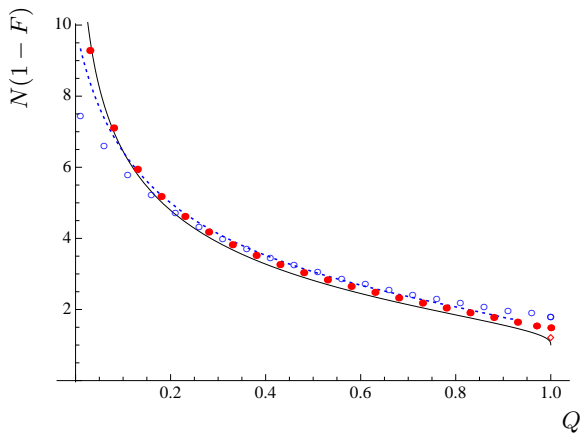}}
\put (-0.2,9.){\rotatebox{90}{$N(1-F)$}}
\put (14.5,.5){$Q$}
\end{picture}
\caption{Plot of $n S_{\rm min}[=N(1-F)]$ vs. $Q$. The solid black line is the leading asymptotic expression in Eq.~(\ref{ebc23.05.13-6}). The dashed line is the curve $(Q,n S_{\rm min})$ given by Eqs.~(\ref{ebc22.05.13-5}) and~(\ref{ebc22.05.13-3}) for~$n=20$. The blue empty (red filled) circles are numerical results for $n=20$ ($n=90$). The empty diamond is also a numerical result for $Q=1$ and $n=1000$.}\label{TvsQ} 
\end{figure}
\end{center}

For $0<Q<1$ ($\lambda>1$) and large $n$, the very same argument we used for large abstention shows that $\varphi(t)$ will be peaked away from $t=0$, at some value close to the boundary of the coincidence set. We can thus Taylor expand the term $2n/t$ in~(\ref{edframes}) around $t=\alpha$ to sub-leading order. The differential equation becomes
%
%
\begin{equation}
{d^2\over d t^2}\varphi+{2n\over \alpha^2}t\,\varphi+\left(\omega^2-{4n\over\alpha}\right)\varphi=\sigma,
\label{ebc21.05.13-4}
\end{equation}
%
whose solution in ${\mathscr C}^c$ (where $\sigma=0$) is
%
\begin{equation}
\varphi(t)=C \; {\rm Ai}\left[
{4\alpha n-\alpha^2\omega^2-2n t \over (2\alpha n)^{2/3}}
\right].
\end{equation}
Here, we have used the weaker boundary condition $\lim_{t\to-\infty}\varphi(t)=0$, and $C$ is determined in terms of the remaining free parameters $\alpha$ and $\omega$ by imposing continuity at the boundary of the coincidence set: $\varphi(\alpha)=\lambda\psi(\alpha)$.
%
%
This combined with continuity of the first derivative implies $\varphi(\alpha)/\varphi'(\alpha)=\psi(\alpha)/\psi'(\alpha)$, thus
\begin{equation}
{\alpha^{2/3}
 {\rm Ai}\left[{\alpha(2n-\alpha\omega^2)\over(2\alpha n)^{2/3}}\right]
\over(2n)^{1/3} {\rm Ai}'\left[{\alpha(2n-\alpha\omega^2)\over(2\alpha n)^{2/3}}\right]}
=1-\alpha. 
\label{ebc22.05.13-1}
\end{equation}
By inspection, we see that in order for this expression to make sense for asymptotically large $n$, the Lagrange multiplier $\omega$
must be of the form
\begin{equation}
\omega^2={2n\over\alpha}-(2\alpha n)^{2/3}{\gamma'_1\over\alpha^2}+\epsilon(n)\, n^{1/3}
\equiv\omega_0^2+O(n^{1/3})
 ,
\label{ebc23.05.13-3}
\end{equation}
with $\epsilon(n)=o(n^0)$ and $\gamma'_1$ being the first zero of the ${\rm Ai}'$ function ($\gamma'_1\approx-1.0188$).
To compute $\epsilon(n)$, we assume it has an asymptotic series expansion in inverse powers of~$n^{1/3}$ and plug it into~(\ref{ebc22.05.13-1}). We then obtain the coefficients of the resulting series recursively.
At leading order we have $\epsilon(n)=-(2\alpha)^{1/3}/[\alpha(1-\alpha)\gamma_1']$.
There is however an additional order $n^{1/3}$ contribution to $\omega^2$ coming from the next (quadratic) order 
in the Taylor expansion of $2n/t$ in~(\ref{edframes}). It can be computed using perturbation theory. Namely, as $\delta\omega^2=(2n/\alpha^3)\int_{-\infty}^\alpha dt \,(\alpha-t)^2\varphi^2$ [in this expression $\varphi$ is assumed to be normalized to one in~$(-\infty,\alpha)]$.
Combining the two order $n^{1/3}$ contributions one has
\begin{equation}
\omega^2\!\!=\!\omega_0^2+{(8\gamma_1'{}^3\!\!-\!3)\!-\!4\alpha(2\gamma_1'{}^3\!\!+\!3)\over15\alpha^2(1-\alpha)\gamma'_1}(2\alpha n)^{1/3}\!+O(n^0) ,
\label{ebc23.05.13-1}
\end{equation}
where $\omega^2_0$ is defined in~(\ref{ebc23.05.13-3}). This equation
gives $\omega^2$ as an explicit function of $\alpha$.
The rate of abstention (equivalently, $\lambda$) can also be expressed as a function of $\alpha$ by imposing normalization to the solution of~(\ref{ebc21.05.13-4}) in the whole interval $(-\infty,1]$, i.e., 
$\int_{-\infty}^1dt\;\varphi^2(t)=1$. 
One has
%
%
\begin{eqnarray}
&&\kern-2em\bar Q=\!{1\over\lambda^2}\!=\!(1-\alpha)^3\nonumber\\
&&\kern-2em\phantom{\bar Q\!=\!{1\over\lambda^2}}
\!-\!
3\alpha(1\!-\!\alpha)^2\!\!\left[\!\left(\!1\!-\!{\alpha\omega^2\over2n}\right)\!-\!{\alpha\over2n(1\!-\!\alpha)^2}\right] .
\label{ebc22.05.13-5}
\end{eqnarray}
%
Using~(\ref{action}) once again, we obtain 
%
%
%
\begin{eqnarray}
S_{\rm min}&=&
-{3\lambda^2\over n}\left(\log\alpha+2-2\alpha-{1-\alpha^2\over2}\right)
\nonumber\\ &+&
{\omega^2\over2n^2}\left[1-\lambda^2(1-\alpha)^3\right]
.\label{ebc22.05.13-3}
\end{eqnarray}
%
Eqs.~(\ref{ebc22.05.13-5}) and ~(\ref{ebc22.05.13-3}), define the curve $(Q,S_{\rm min})$ in terms of $\alpha$, which we view as a free parameter that takes values in the range $0<\alpha<1$. This curve, which is accurate up to order~$n^{-4/3}$, is plotted in Fig.~\ref{TvsQ} (dashed line) for~$n=20$. In the same figure, we also plot the asymptotic  (leading) contribution alone (solid line) and some numerical optimization results for $n=20$ (empty blue circles) and~$n=90$ (filled red circles). We see that~$n=20$ is still not quite in the asymptotic regime,  and that the sub-leading corrections play a significant role,  improving the agreement to almost perfect for central values of $Q$.

At leading order, $S_{\rm min}$ can be easily written as an explicitly function of $Q$, since only the leading term in~(\ref{ebc22.05.13-5}) contributes and we have
$\alpha=1-\bar Q{}^{1/3} $.
Substituting this in the first line of~(\ref{ebc22.05.13-3}), we obtain
\begin{equation}
S_{\rm min}\!=-{3\over 2n \bar Q}\!\left[2\log(1-\bar Q{}^{1/3})+2\bar Q{}^{1/3}\!+\bar Q{}^{2/3}\right].
\label{ebc23.05.13-6}
\end{equation}
Interestingly, the corrections to this result can be shown to be of order~$n^{-5/3}$, whereas the implicit form given by~(\ref{ebc22.05.13-5}) and~(\ref{ebc22.05.13-3}) has non-zero contributions of order~$n^{-4/3}$. 
In the limit $Q\rightarrow 1$, Eq.~(\ref{ebc23.05.13-6}) yields the leading order in~\eqref{ebc07.06.13-1}, but the slope of $S_{\rm min}(Q)$ becomes vertical at $Q=1$ (see solid line in Fig.~\ref{TvsQ}). 
At this point our asymptotic approximation breaks down ---as can be seen by noticing that the higher order terms, e.g., Eq.~(\ref{ebc23.05.13-1}), diverge as some negative power of~$1-\alpha\approx \bar Q{}^{1/3}$--- and the numerical results approach the leading asymptotic curve very slowly. This is apparent from Fig.~\ref{TvsQ}, where an extra point (empty diamond), corresponding to a numerical result for $n=1000$, has been added to 
further emphasize this behavior.
%

At the other end, for $Q\to0$, Eq.~(\ref{ebc23.05.13-6}) diverges. That should not come as a surprise, since, as mentioned above, for zero abstention the error scales as $(1/n)\log n$. This also explains why the agreement with the numerical results (circles) in Fig.~\ref{TvsQ} worsens as $Q$ becomes very small.

\section{Conclusions}
We have studied the effect of abstention, or post-selection, in parameter estimation.
In some cases, such as that of $N$ parallel spins encoding a spatial direction, abstention does not provide any enhancement of the estimation precision. However, generically post-selection do have a significant effect
, even asymptotically. 

The problem of finding the optimal protocol with abstention can be rephrased 
as that of optimizing the probabilistic map that transforms the family of input states into a new family that yields a higher estimation fidelity. 
The optimization is first formulated as a SDP problem, which immediately renders it numerically solvable.
Most importantly, we have also presented a method for computing the fidelity and the form of the transformed states as a function of the abstention rate $Q$ for asymptotically large samples. 
This method relies on mapping our optimization problem to a mechanical problem defined through an effective Lagrangian (action) where the input state plays the role of a moving constraint. Solving the corresponding equations of motion returns the optimal fidelity for a fixed abstention rate $Q$, and the corresponding optimal POVM.
We have given the general form of this Lagrangian for the relevant problems of phase, direction and Cartesian frame estimation, 
and thereby cleared the road for finding analytical optimal solutions for arbitrary input states. We would like to emphasize that this is a significant development, since even in the standard approach to estimation, without abstention, analytical asymptotic expressions were only known in few cases. 

For phase and direction estimation we have  illustrated our method for two  types of input states. 
We have first studied states proportional to 
the (rotated) seed vector of the respective optimal POVM in standard parameter estimation, 
and then moved into product states of identically prepared qubits, polarized on the equatorial plane  (phase), and into products of pairs of antiparallel spins (direction). The rate at which the fidelity approaches one establishes two distinct regimes: In the first regime the rate is proportional to $N^{-1}$ (the so-called shot noise limit) and the abstention can change the proportionality constant up to a factor of two. This means that in a given setup an experimentalist would attain the same gain in fidelity by cranking up the number of copies than she would by allowing for some degree of  abstention. The second regime is much more dramatic: the fidelity approaches one as $N^{-2}$, thus attaining the Heisenberg limit. The abstention rate that separates the two regimes depends on the input states under consideration.
For 
 input states proportional to the rotated POVM seeds, which have a very broad distribution in the relevant quantum number but provide a shot-noise limited fidelity in standard estimation (without abstention), the slightest 
abstention rate, $Q> 0$, is enough to unlock the good encoding properties of these states and reach the Heisenberg, $N^{-2}$, regime. Product states can also reach this enhanced regime, but in this case the abstention rate needs to get exponentially close to one.
In contrast to the previous case studied in~\cite{gendra}, where the action of the POVM can be understood as a filtering of subspaces preceding the optimal canonical measurement, here the POVM plays a more active role and modifies in a non-trivial way the coherences in the states.
The benefits of abstention are also more visible here than in Ref.~\cite{gendra}, where
 an exponentially small acceptance rate was required to change the coefficient of the shot noise term~$N^{-1}$, and  the Heisenberg regime was not attainable at all.

Cartesian frame estimation  has been shown to be formally equivalent to phase estimation in the asymptotic regime of many spins, provided one can entangle the magnetic number $m$ with the quantum number that labels the degeneracy of the total angular momentum representations. 
%
In addition, we have studied frame estimation with systems where no such degeneracy exists (such as Rydberg atoms) or cannot be exploited. The method is illustrated for a simple input state  where the amplitudes of the different angular momentum eigenstates are linearly decreasing  with $j$. In this case, even a tiny amount of abstention triggers a change in the averaged error scaling, from $(1/N)\log N$ to $1/N$, which is the fastest decrease one can attain in this scenario. Increasing the abstention rate further reduces the scaling-law coefficient down to almost its minimum value. 



Recently~\cite{tutorial}, there has been revamped interest in weak measurements~\cite{vaidman}, with particular emphasis on quantum metrology ~\cite{kwiat,tanaka}. The protocol of state estimation with abstention presented here and weak measurements are both instances of post-selection. Our~framework does not assume any specific 
realisation of the measurements, therefore the bounds derived here also apply to a weak measurement set-up. Note, however, that
most of the work on weak-measurement metrology follow a point-wise approach to estimation, as opposed to the Bayesian approach followed here (see however~\cite{tanaka}).  The analysis of abstention in a point-wise approach together with the important extension of our methods to mixed states will be presented in~\cite{abst-mixed}. 
Finally, we note that very recently a similar use of abstention has been applied to other quantum processing tasks, such as quantum cloning (or replication)~\cite{replication} achieving also enhanced efficiency.

 

\begin{acknowledgments}
We acknowledge financial support from ERDF: European Regional Development Fund. This research was supported by
 the Spanish MICINN, through contract FIS2008-01236 and the Generalitat de
Catalunya CIRIT, contract  2009SGR-0985. We thank Giulio Chiribella for inspiring discussions at the early stages of this work and Madalin Guta for pointing out its  relation to weak measurements.
\end{acknowledgments}

\end{document}